\documentclass{elsart}
\usepackage{graphicx}
\graphicspath{{figs/}}
\usepackage{amssymb}

\begin{document}

\begin{frontmatter}
\title{The MiniBooNE Detector}
\collaboration{The MiniBooNE Collaboration}
\author[columbia]{A.~A. Aguilar-Arevalo},
\author[yale]{C.~E.~Anderson},
\author[fnal]{L.~M.~Bartoszek},
\author[princeton]{A.~O.~Bazarko},
\author[fnal]{S.~J.~Brice},
\author[fnal]{B.~C.~Brown},
\author[columbia]{L.~Bugel},
\author[umich]{J.~Cao},
\author[columbia]{L.~Coney},
\author[columbia]{J.~M.~Conrad},
\author[indiana]{D.~C.~Cox},
\author[yale]{A.~Curioni},
\author[columbia]{Z.~Djurcic},
\author[fnal]{D.~A.~Finley},
\author[yale]{B.~T.~Fleming},
\author[fnal]{R.~Ford},
\author[fnal]{F.~G.~Garcia},
\author[lanl]{G.~T.~Garvey},
\author[lanl,fnal]{C.~Green},
\author[indiana,lanl]{J.~A.~Green},
\author[colorado]{T.~L.~Hart},
\author[lanl,cinci]{E.~Hawker},
\author[lsu]{R.~Imlay},
\author[cinci]{R.~A. ~Johnson},
\author[columbia]{G.~Karagiorgi},
\author[fnal]{P.~Kasper},
\author[indiana]{T.~Katori},
\author[fnal]{T.~Kobilarcik},
\author[fnal]{I.~Kourbanis},
\author[bucknell]{S.~Koutsoliotas},
\author[princeton]{E.~M.~Laird},
\author[yale]{S.~K.~Linden},
\author[vtech]{J.~M.~Link},
\author[umich]{Y.~Liu},
\author[bama]{Y.~Liu},
\author[lanl]{W.~C.~Louis},
\author[columbia]{K.~B.~M.~Mahn},
\author[fnal]{W.~Marsh},
\author[fnal]{ P.~S.~Martin},
\author[lanl]{G.~McGregor},
\author[lsu]{W.~Metcalf},
\author[indiana]{ H.-O.~Meyer},
\author[princeton]{P.~D.~Meyers},
\author[fnal]{F.~Mills},
\author[lanl]{G.~B.~Mills},
\author[columbia]{J.~Monroe},
\author[fnal]{C.~D.~Moore},
\author[colorado]{R.~H.~Nelson},
\author[columbia]{V.~T.~Nguyen},
\author[marys]{P.~Nienaber},
\author[lsu]{J.~A.~Nowak},
\author[lsu]{S.~Ouedraogo},
\author[princeton]{R.~B.~Patterson},
\author[bama]{D.~Perevalov},
\author[indiana]{C.~C.~Polly},
\author[fnal]{E.~Prebys},
\author[cinci]{J.~L.~Raaf},
\author[lanl,florida]{H.~Ray},
\author[umich]{B.~P.~Roe},
\author[fnal]{A.~D.~Russell},
\author[lanl]{V.~Sandberg},
\author[princeton]{ W.~Sands},
\author[lanl]{R.~Schirato},
\author[lsu]{G.~Schofield},
\author[columbia]{D.~Schmitz},
\author[columbia]{M.~H.~Shaevitz},
\author[princeton]{F.~C.~Shoemaker},
\author[embry]{D.~Smith},
\author[yale]{M.~Soderberg},
\author[columbia]{M.~Sorel}\footnotemark[1],
\author[fnal]{P.~Spentzouris},
\author[bama]{I.~Stancu},
\author[fnal]{R.~J.~Stefanski},
\author[lsu]{M.~Sung},
\author[princeton]{H.~A.~Tanaka},
\author[indiana]{R.~Tayloe},
\author[colorado]{M.~Tzanov},
\author[lanl]{R.~Van~de~Water},
\author[lsu]{M.~O.~Wascko}\footnotemark[2],
\author[lanl]{D.~H.~White},
\author[colorado]{M.~J.~Wilking},
\author[umich]{H.~J.~Yang},
\author[columbia,lanl]{G.~P.~Zeller},
\author[colorado]{E.~D.~Zimmerman}
\address[bama]{University of Alabama, Tuscaloosa, AL 35487}
\address[bucknell]{Bucknell University, Lewisburg, PA 17837}
\address[cinci]{University of Cincinnati, Cincinnati, OH 45221}
\address[colorado]{University of Colorado, Boulder, CO 80309}
\address[columbia]{Columbia University, New York, NY 10027}
\address[embry]{Embry-Riddle Aeronautical University, Prescott, AZ 86301}
\address[fnal]{Fermi National Accelerator Laboratory, Batavia, IL 60510}
\address[florida]{University of Florida, Gainesville, FL 32611}
\address[indiana]{Indiana University, Bloomington, IN 47405}
\address[lanl]{Los Alamos National Laboratory, Los Alamos, NM 87545}
\address[lsu]{Louisiana State University, Baton Rouge, LA 70803}
\address[umich]{University of Michigan, Ann Arbor, MI 48109}
\address[princeton]{Princeton University, Princeton, NJ 08544}
\address[marys]{Saint Mary's University of Minnesota, Winona, MN 55987}
\address[vtech]{Virginia Polytechnic Institute \& State University, Blacksburg, VA 24061}
\address[yale]{Yale University, New Haven, CT 06520}
\footnotetext[1]{Present address: IFIC, Universidad de Valencia and CSIC, Valencia 46071, Spain}
\footnotetext[2]{Present address: Imperial College London, London, SW7 2AZ, UK}

\begin{abstract}
The MiniBooNE neutrino detector was designed and built to look for $\nu_\mu
\rightarrow \nu_e$ oscillations in the $(\sin^{2}2\theta,\Delta m^2)$ parameter
space region where the LSND experiment reported a signal.
The MiniBooNE experiment used a beam energy and baseline that were an order of
magnitude larger than those of LSND so that the backgrounds and systematic
errors would be completely different. This paper provides a detailed
description of the design, function, and performance of the MiniBooNE detector.

\end{abstract}
\end{frontmatter}

\section{Introduction}
\subsection{Physics Overview}

The MiniBooNE experiment at Fermilab  was proposed to test the evidence for
neutrino  oscillations from the LSND experiment at Los Alamos \cite{lsnd}.  The
LSND experiment observed more $\bar \nu_e$ candidate events than expected from
background. If the excess is interpreted  as being due to $\bar \nu_\mu
\rightarrow \bar \nu_e$ oscillations, then the most favored oscillation region
is a band in $\Delta m^2$ stretching from $\sim$0.2~eV$^2$ to $\sim$2~eV$^2$.
The MiniBooNE experiment was designed to search for $\nu_\mu \rightarrow \nu_e$
and  $\bar \nu_\mu \rightarrow \bar \nu_e$ oscillations with approximately the
same $L/E \simeq 1$ value as LSND, where $L$ is the neutrino travel distance from
the source to the detector in meters and $E$ is the neutrino energy in MeV.
Whereas the LSND neutrino beam travelled a distance of 30~m with a typical
energy of 30~MeV, the MiniBooNE neutrino beam traveled 500~m and had a typical
energy of 500~MeV. With neutrino energies an order of magnitude higher, the
MiniBooNE backgrounds and systematic errors are completely different from those
of LSND. MiniBooNE, therefore, constitutes an independent check of the LSND
evidence for neutrino oscillations at the $\sim$1~eV$^2$ mass scale.
          
\subsection{Physics Driven Parameters}

In order to search effectively for  $\nu_\mu \rightarrow \nu_e$ and  $\bar
\nu_\mu \rightarrow \bar \nu_e$ oscillations, the MiniBooNE detector needed to
satisfy certain requirements. First, the detector required a target mass of
$\sim$1~kton in order to  generate $\sim$1000 neutrino oscillation events
for $10^{21}$ protons on target. Second, the detector needed to provide
excellent discrimination between $\nu_\mu$ and $\nu_e$ induced events. The scale
is set by the LSND neutrino oscillation probability of $\sim$0.26\%.
(The intrinsic $\nu_e$ background in MiniBooNE is $\sim$0.5\%.) Third, the
detector had to have a completely active volume with no dead regions. This was
necessary in order to contain neutral-current $\pi^0 \rightarrow \gamma \gamma$
events, which would constitute a large background if one of the $\gamma$'s 
escaped detection. Fourth, the detector needed to have a 4$\pi$ veto to reject
cosmic ray events,  neutrino interactions that occur outside the detector, and
neutrino events with  tracks that escape the fiducial volume. Liquid Cherenkov
detectors have no dead regions, have an easily configured veto region, and,
thanks to modern computers, have excellent particle identification. A liquid
Cherenkov detector is an economical choice that meets all of these
requirements.

\subsection{Overall Design Considerations and Constraints}

Mineral oil was chosen  instead of water as the liquid for the MiniBooNE
detector for  several reasons. First, mineral oil has an index of refraction
$n=1.47$, which is considerably higher than the $n=1.33$ index of refraction
for water. This higher index of refraction, together with a lower density than
water (0.85~gm/cm$^3$ instead of 1.00~gm/cm$^3$), means that electrons  produce considerably
more Cherenkov light in mineral oil than in water. Furthermore, the lower
velocity of light in mineral oil improves the event position reconstruction.
Second, mineral oil allows the detection of lower-energy muons,
pions, and protons than in water due to the lower Cherenkov threshold and the
presence of scintillation light in pure mineral oil. This is used for
background rejection and for measuring backgrounds down to lower energies.
Third, mineral oil has less multiple scattering than water and a smaller
$\mu^-$ capture rate, 8\% compared to 20\% in water. The smaller $\mu^-$
capture rate increases the efficiency of the identification of charged-current
reactions using the Michel electron tag from muon decay. Mineral oil has
the additional advantage that one can safely immerse electronic components in
it. The downside of mineral oil is that it requires a much more complicated
optical model to describe the generation and transmission of light through the
medium (see Sec.~\ref{sec:oil_prop}).

\begin{figure}
\centerline{\includegraphics[width=3.0in]{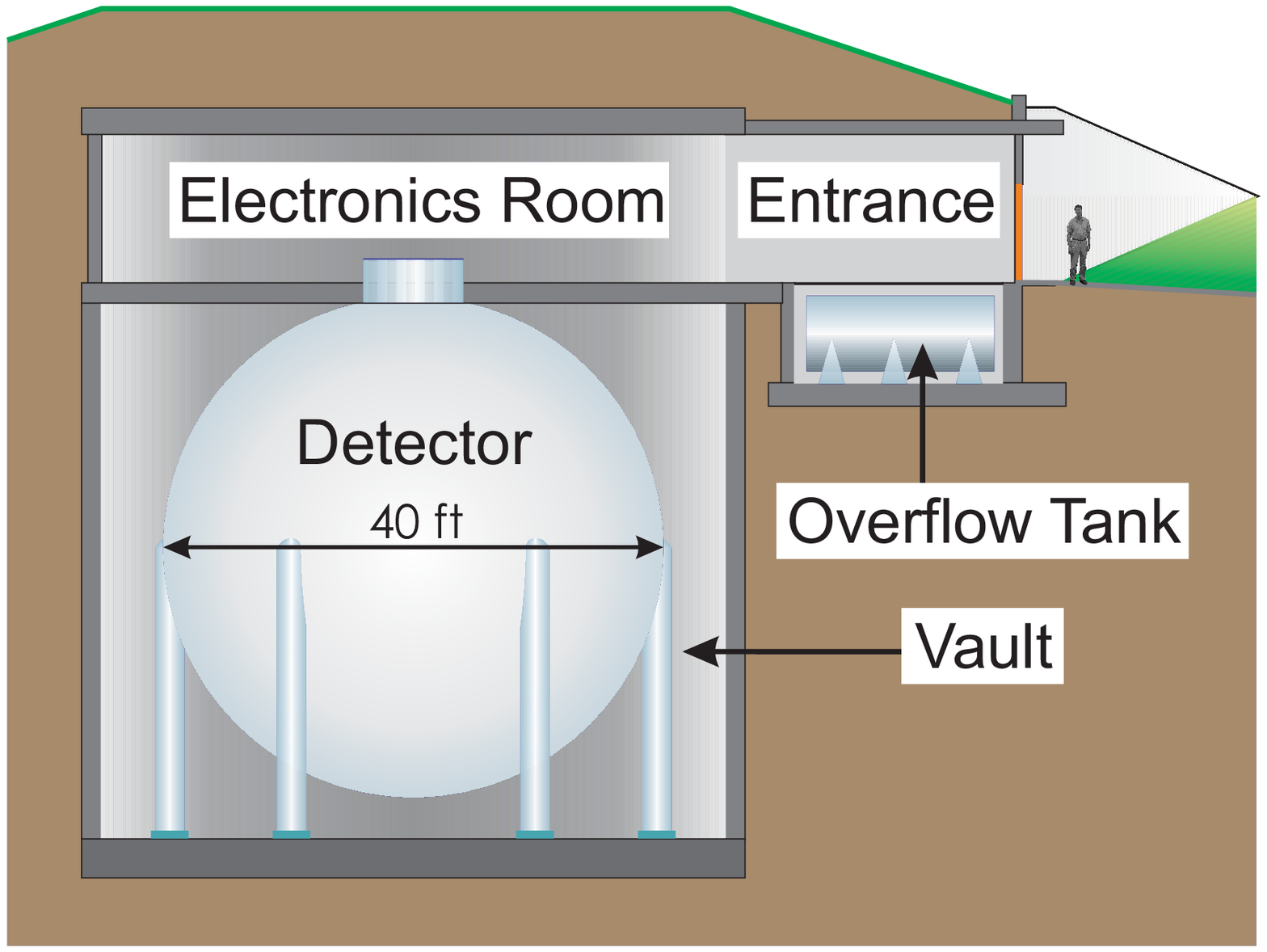}
\hfil \includegraphics[width=2.0in]{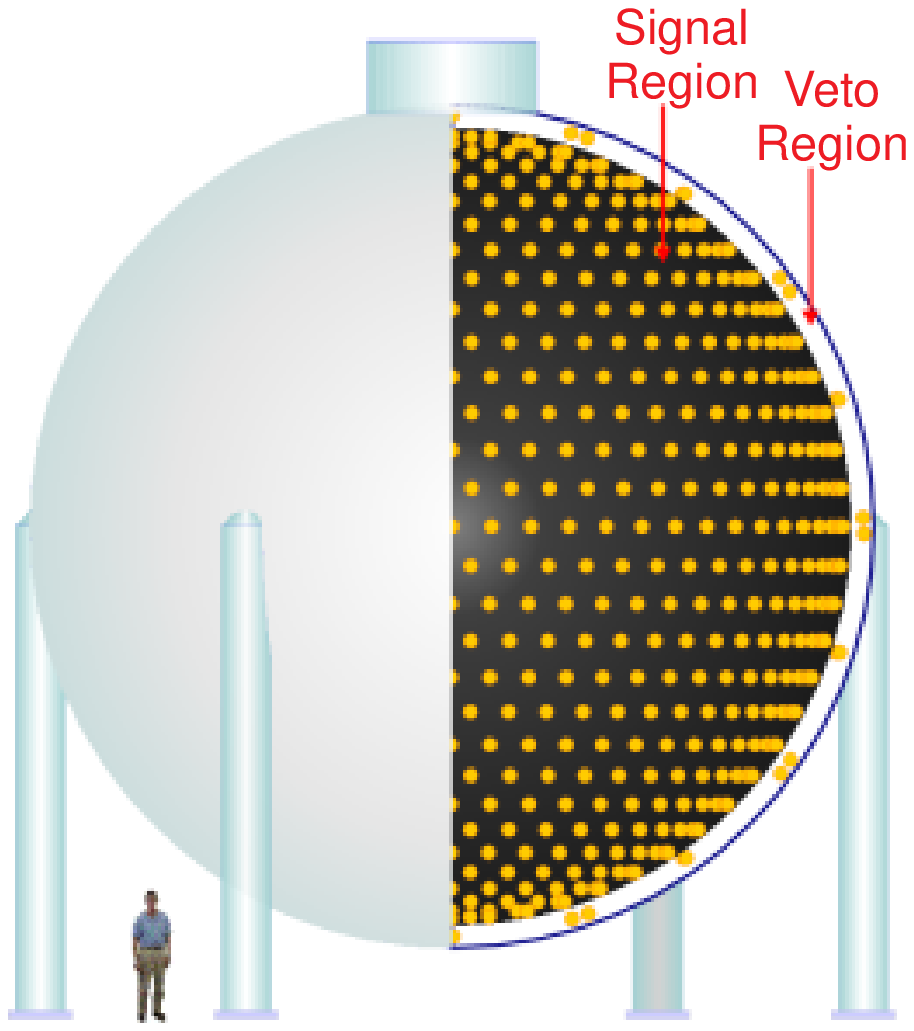}}
\caption{The MiniBooNE detector enclosure (left) and a
cut-away drawing (right) of the detector showing the distribution of PMT's in the
signal and veto regions.}
\label{boone_tank}
\end{figure}

As the photomultiplier tube (PMT) 
coverage for a liquid Cherenkov  detector is proportional to
the detector surface area, a spherical  tank was chosen to maximize the ratio
of volume to surface area. Furthermore, a spherical geometry has no inside
edges which is beneficial for the  event reconstruction.  The detector (see
Fig.~\ref{boone_tank}) is a spherical tank of diameter 12.2~m (40~ft), which is
filled with 818 tons of mineral oil. An opaque barrier divides the volume into
an inside main detector region and an outside veto region and supports the
PMT's viewing the main detector region.

In order to reduce the detector cost, the collaboration chose to reuse the LSND
phototubes ($\sim$1220) and electronics ($\sim$1600 channels). An additional 330
phototubes were purchased in order to obtain a total phototube channel count of
1520 after  rejection of the poorest tubes.
The allocation of PMT's in the main tank and veto and the thickness
of the veto region were determined by physics considerations and were
arrived at using Monte Carlo simulations of signal and background events.
The Monte Carlo studies used a full GEANT
simulation, including tracking of individual Cherenkov and scintillation
photons, with wavelength-dependent absorption, reflection, and detection
efficiencies.  Analysis of events in the main tank indicated that at least
10\% photocathode coverage (defined by treating the photocathodes as flat
disks with diameter equal to the PMT diameter)
was needed to provide the required particle identification quality.
When tuned to the secondary requirement that veto and main tank channels not 
be mixed in the same electronics crate, a final number of 1280 tank PMT's
resulted.  Calculated with the final radial position, this allocation has a 
photocathode coverage of 11.3\%.

This allowed 240 veto PMT's, and the issue was whether the veto region would
yield sufficient light to reject muons from cosmic rays and beam-neutrino
interactions outside the tank with an efficiency $>$99\%, the design goal. Monte
Carlo studies, including estimates of the albedo of the painted
surfaces in the veto (see Sec.~\ref{sec:paint}),  indicated that the light yield
with a 35~cm veto thickness was high enough to allow reasonable thresholds in
the face of noise and an estimate of light leaked from the main
tank.  Separate  calculations showed that loss of signal due to  muon tracks
hiding in the struts and cables that penetrate the veto region was negligible.
The actual efficiency for rejecting cosmic ray muons was measured to be 99.99\%
(see Sec.~\ref{sec:neutrinos}). The 35~cm veto thickness puts the optical
barrier at a radius of 574.6~cm.

\subsection{Experiment Layout}
Protons with 8~GeV kinetic energy are extracted from the Fermilab Booster and
transported to the MiniBooNE target hall which contains a beryllium target
within a magnetic focusing horn. The target and horn are followed by a pion
decay volume at the end of which is a 3.8~m thick steel and concrete
beam dump. The distance from the center of the target to the front face of the
dump is 50~m.

The neutrino beam~\cite{flux_paper} that is produced in the decay volume passes through the dump
plus 474~m of earth before reaching the MiniBooNE detector vault. This ensures
that neutrinos are the only beam products that can reach the detector. The total
distance from the upstream face of the target to the center of the detector is
541~m.

The detector tank, as shown in Fig.~\ref{boone_tank}, sits below ground level
inside a 13.7~m (45~ft) diameter cylindrical vault with a room above that
houses electronics and utilities. The vault not only provides access to the
tank's exterior plumbing, but also acts as secondary containment for the mineral
oil. The entire structure is covered by at least 3~m of dirt which provides
shielding against cosmic ray backgrounds and makes it easier to keep the
detector at a constant temperature.

\section{Photomultiplier Tubes}
\subsection{PMT Specification}

\begin{figure}
\centerline{\includegraphics[width=6.5in]{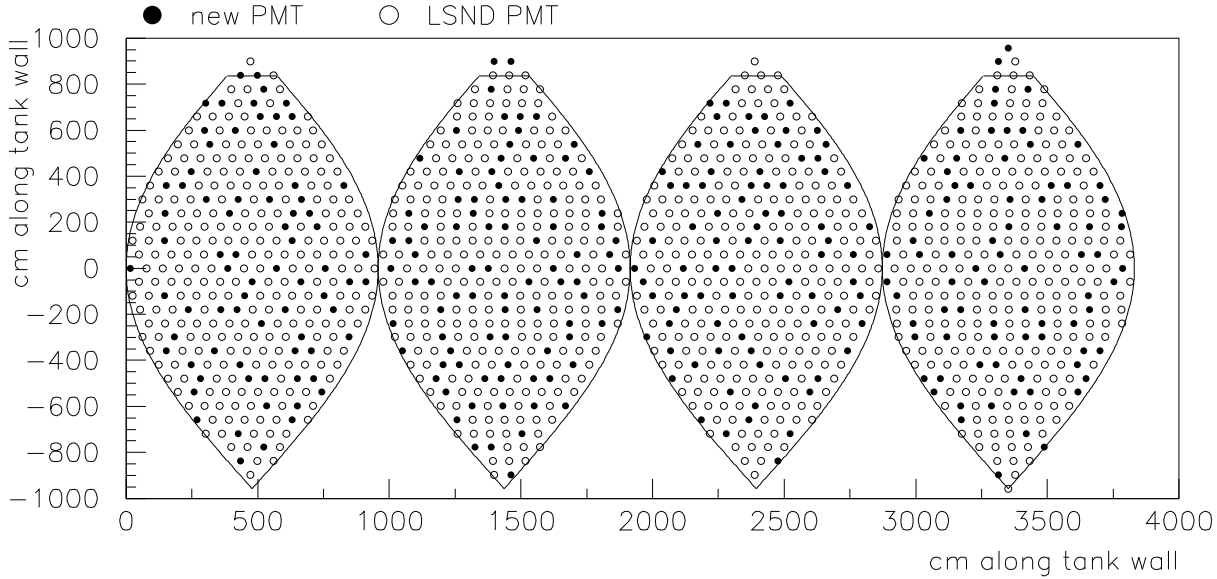}}
\caption{Main phototube layout.  PMT's are not drawn to scale.}
\label{fig:pmt_map}
\end{figure}

The PMT's used in MiniBooNE
were 8~inch Hamamatsu tubes of two types: 1198 were R1408's recycled from LSND,
and 322 were newer R5912 tubes. As detailed in Ref.~\cite{bonnie}, all the PMT's
were tested before installation to determine their relevant properties: gain,
dark noise rate, charge resolution, timing resolution, and double pulsing rate.
As the new tubes have better time and charge  resolution, all were used in the
main tank, distributed randomly. The older tubes with the lowest dark noise rate
were used in the veto and the rest completed the main tank array. The array of
tubes deployed in the main tank is shown in Fig.~\ref{fig:pmt_map}.

These PMT's have a wavelength dependent efficiency that, as shown in
Fig.~\ref{fig:albedo}, peaks at 390 nm, falling to half its maximum at 315 and 
490 nm~\cite{hamamatsu}.  The MiniBooNE PMT's are operated with
$\sim$2000~V (positive) on the dynode chain, resulting in a gain of $\sim1.6
\times 10^7$. The intrinsic time resolution of the PMT's is $\sim$1~ns, and the
intrinsic charge resolution is $\sim$15\% at
1~p.e~\cite{Overview::Gladstone:2006}. Although the time and charge measurements
are smeared by the data acquisition electronics, the dominant
contribution to the resolutions results from the intrinsic PMT properties.

\subsection{PMT preparation}
\label{sec:pmt_prep}

After the PMT's had their bases attached, were tested, and were assigned
to a location in the tank, they were prepared for installation.

\begin{description}
\item[$\bullet$] Each tube was washed in a mild solution of detergent and distilled
water, rinsed in clean distilled water, and then allowed to dry for 24 hours.
The main 
purpose was to remove the scintillator-doped oil residue from LSND.
\item[$\bullet$] New tubes were mounted in stands (see
Sec.~\ref{sec:pss_struct}).  Old tubes destined for the main tank were
re-mounted in stands.
\item[$\bullet$] The tube was dipped in black Master Bond EP21LV encapsulant
from the base up to the bottom of the globe.  This is to protect the oil from
the components of the base and vice versa.
Old tubes were already coated in black Hysol.  This was found to contaminate
mineral oil (Sec.~\ref{sec:oil_compat}), so the old tubes were also
dipped in Master Bond.
\item[$\bullet$] Veto tubes were removed from their stands and mounted in
veto mounts.
\item[$\bullet$] Main tubes had the centering clips attached around the tube
neck. When LSND was opened to harvest the phototubes, it was found that about
100 tubes had floated out of their stands. Consequently, a stripe of Master Bond
was run down the neck of the tube  over the silicone rubber cushion on each
centering clip.  This attachment  served as a mechanical backup in case the
metal band and foam tape at the tube's equator slipped.
\end{description}

This work was done in a cleanroom.  It was found that maintaining low 
humidity was necessary for the proper curing of the Master Bond.

\section{Mineral Oil}
\subsection{Oil Selection and Delivery}
The MiniBooNE detector tank holds $9.5\times10^5$ 
liters of mineral oil.  Because of the large path length that an optical photon
must traverse in the oil, a large extinction length (attenuation of a beam due to 
all causes including scattering) was the most important
criterion for the mineral oil chosen for MiniBooNE.  
An extinction length of greater than
20~m for 420~nm light was required in order to lose no more than 25 percent of
the light generated by a neutrino interaction occurring in the center of the
detector.  Light transmission
throughout the wavelength range of 320~nm to 600~nm was also considered in selecting
the oil.
The oil was also required to have a large index of
refraction, a small dispersion over the wavelength range 320~nm to 600~nm, low
reactivity with materials in the detector, low levels of scintillation light,
and low viscosity~\cite{Raaf:2002mt}.

Ten different oils from six different vendors were tested. Based on the
above criteria,  Marcol 7 Light Mineral Oil (Industrial NF grade), an Exxon/Mobil product
manufactured by Penreco was selected for MiniBooNE. 
It was certified by the manufacturer to have a
specific gravity between 0.76 and 0.87~gm/cm$^3$ as measured via the American Society
for Testing and Materials (ASTM) D~4052 or ASTM~D~1298, a viscosity less than
34.5 cSt at $40~^\circ{\mathrm{C}}$ as measured via ASTM~D~445, and a color
greater than or equal to 30 Saybolt Color units as measured via ASTM~D~156.
While a more dense oil would have provided more interactions in the detector,
the need for recirculation imposed an upper limit on its viscosity and an
implicit maximum density. 

The Marcol 7 mineral oil was delivered to the Fermilab railhead in food-grade
railcar tankers that were cleaned and dried before filling. Prior to shipping,
the manufacturer was required to draw a sample from the bottom of each railcar
and perform the following tests using the methods shown in parentheses: specific
gravity (ASTM~D~4052 or D~1298), color (ASTM~D~156), cloud point (ASTM~D~2500),
pour point (ASTM~D~97), kinematic viscosity (ASTM~D~445), and flash point
(ASTM~D~92). Upon arrival at Fermilab, but before accepting delivery of each
railcar, a sample of the delivered oil was tested to ensure that its extinction
length and light transmission curve agreed with earlier measurements made on the
samples submitted during the oil selection process. Once the oil was verified to
be Marcol 7 with an acceptable extinction length, it was offloaded from the
railcar via clean plumbing to food grade tanker trucks that had been visually
inspected for dirt or damage. A vent cover and dry air line was installed on the
railcar before offloading in order to avoid contamination of the oil. Transfer
from the tanker trucks to the detector also made use of clean plumbing, vent
covers, and dry air to avoid contamination.

\subsection{Oil Properties}
\label{sec:oil_prop}

Further measurements were performed on the mineral oil after its delivery
to better determine its properties in order to model optical photon
transport in the detector simulation. 
Some of these measurements are presented in Table~\ref{tab:oilprop}.

\begin{table}
\centerline{\begin{tabular}{|l|l|} \hline
Density & $0.845\pm0.001~\mathrm{g/cm^{3}}$\\ \hline
Coefficient of volume expansion & $(61\pm 4)\times 10^{-5}$ K$^{-1}$\\ \hline
Refractive index, $n_D$(589.3~nm) & $(1.4684\pm0.0002)\times$\\
           & $\;\left[1-(3.66\pm0.04)\times10^{-4}(T-20\;^\circ\mathrm{C})\right]$\\ \hline
Dispersion ($n_F$(486.1~nm)-$n_C$(656.3~nm)) & $0.0081\pm 0.0003$ \\ \hline
Extinction length (at 460~nm) & 25-40 m (see text)\\ \hline
Rayleigh scattering length (at 442 nm) & $51.7\pm7.0$ m\\ \hline
\end{tabular}}
\caption{Some measured properties of the MiniBooNE Marcol 7 mineral
oil~\cite{Overview::Brown:2004}.}
\label{tab:oilprop}
\end{table}

The density and the coefficient of volume expansion were measured with a 20~ml
pycnometer. The index of refraction and dispersion were measured with an Abbe
Refractometer (Model~WY1A, Xintian Fine Optical Instrument Corp., China) at a
temperature of 20$^\circ\mathrm{C}$. The temperature dependence was also
measured and is given in Table~\ref{tab:oilprop}.

The angular distribution of Rayleigh scattering was measured at 442 and 
532 nm~\cite{Overview::Brown:2004}.
All combinations of incident and scattered polarization with respect to the
scattering plane were measured, allowing isolation of components due to isotropic
and anisotropic density fluctuations.  Absolute measurement of the scattering
rate was accomplished by comparison to the calculable Mie scattering from a 
suspension of 50 nm polystyrene spheres.

\begin{figure}[t,p]
\centerline{\includegraphics[width=5in]{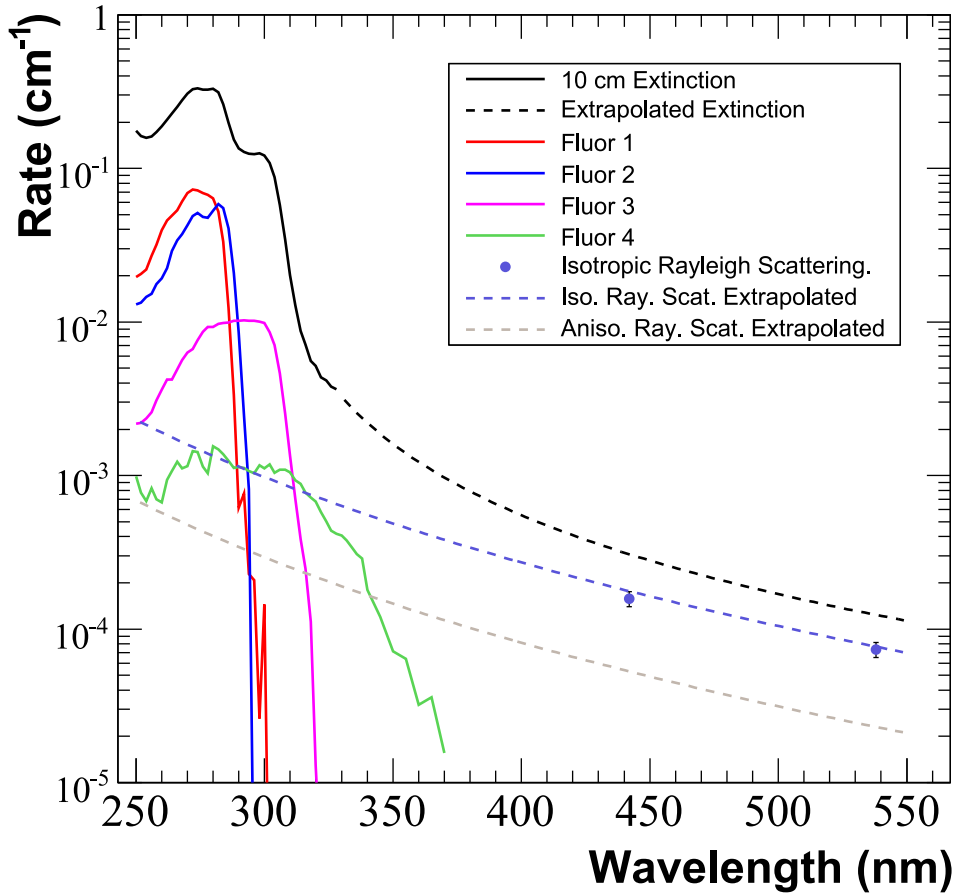} }
\caption{Extinction rate in Marcol 7 mineral oil and some of the processes
contributing to it.  The solid curves and data points are laboratory measurements
on small samples~\cite{Overview::Brown:2004}.  The dashed curves are 
extrapolations after empirical adjustment using MiniBooNE data.
The black curves show the total extinction rate, which is measured directly.
It must include the other measured processes:
fluorescence (four measured fluorophores are shown), Rayleigh scattering (two measured components
shown), and a small contribution from Raman scattering (not shown). 
The remainder  of the extinction is inferred to be absorption.}
\label{fig:extinction}
\end{figure}

Both time-resolved and DC spectrofluorometer measurements were used to characterize the 
fluorescence of the mineral oil~\cite{Overview::Brown:2004}, with lifetimes and
excitation and emission spectra determined for several identified fluorophores.
The measured lifetimes ranged from 1~ns to 33~ns.  These measurements also characterized
a small rate of Raman scattering in the oil.

To obtain the light extinction function for the mineral oil, we performed
several sets of measurements. At wavelengths $\lambda >340$~nm, where the
extinction length exceeds a few meters, the shape of the light transmission
curve as a function of wavelength at a fixed path length and the absolute
transmission of light at 460~nm as a function of path length were measured. The
latter gave an extinction length of $\sim$25~m at 460~nm. A transmission
function of both wavelength and path length was then obtained by scaling the
relative, wavelength-dependent transmission curve to the absolute, 
path-length-dependent curve~\cite{Raaf:2002mt}. Below 340~nm, transmission through 10~cm of
oil was measured in a  spectrophotometer~\cite{Overview::Brown:2004}. 

When the
optical model developed from these measurements was compared to MiniBooNE data
(primarily electrons from at-rest decays of cosmic ray muons), we found that
the data preferred a longer extinction length than the direct measurements at
the longer wavelengths, $\sim$40~m at 460~nm. This difference has not been
resolved. The 40~m extinction length at 460~nm, combined with other parameters
adjusted simultaneously, gives a good representation of MiniBooNE data and is
used in our simulations. 
Various elements from the final optical model are shown in Fig.~\ref{fig:extinction}.
In this model, above 400~nm the extinction is
due chiefly to Rayleigh scattering; below 300~nm it is mostly due to absorption
and fluorescence.

\subsection{Oil compatibility}
\label{sec:oil_compat}

All materials in the tank (PMT's, cables, and support structures) are immersed
in mineral oil, therefore it was necessary to make sure that these materials would not 
contaminate the oil over the duration of the experiment. The danger is that the
contamination could cause changes to the optical properties of the oil that
could adversely affect the performance of the detector or make it
difficult to model. The materials and their potential for contaminating the
mineral oil are:
\begin{itemize}
\item Painted surfaces:  Of particular concern was the leaching of  residual
solvent from incompletely cured paint.  Solvents  contribute additional
scintillating components to the mineral oil.
\item Plastics:   Plastic components, such as cable jackets,  contain
plasticizers, added to a polymer in order to produce a flexible plastic, which
can be the cause of contamination that could affect the oil extinction length. 
\item Metals: The carbon steel tank was painted, as were the aluminum 
components of the phototube support structure.  Nevertheless, some  metal may be
exposed to the oil, through pores and nicks in the paint, and because some
``buried'' surfaces, like the inside of metal tubing, could not be
painted.  It is believed  that metals do not cause contamination of
the mineral oil directly, but their presence acts as a catalyst for oxidation,
particularly at temperatures above $60^\circ$C.  Oxidation decreases the oil
extinction length.
\end{itemize}

Any material proposed for use in the tank was soaked in oil, usually for at least
a  week and often longer, with a sample at room temperature and another at
$66^\circ$C ($150^\circ$F) to speed the aging process.  Each sample was paired
with a control that contained the same oil as the test sample, 
but with no potential contaminants added.
The controls were drawn at the same time as the test samples and treated identically, 
including having 
the same
type of container and same temperature profile.
The optical properties of the
oil were then measured. A spectrophotometer was used to look for changes in
light transmission, and a fluorometer was used to look for changes in
scintillation properties. 

\begin{figure}[t,h,p]
\centerline{\includegraphics[width=5in,bb=23 160 532 655,clip]{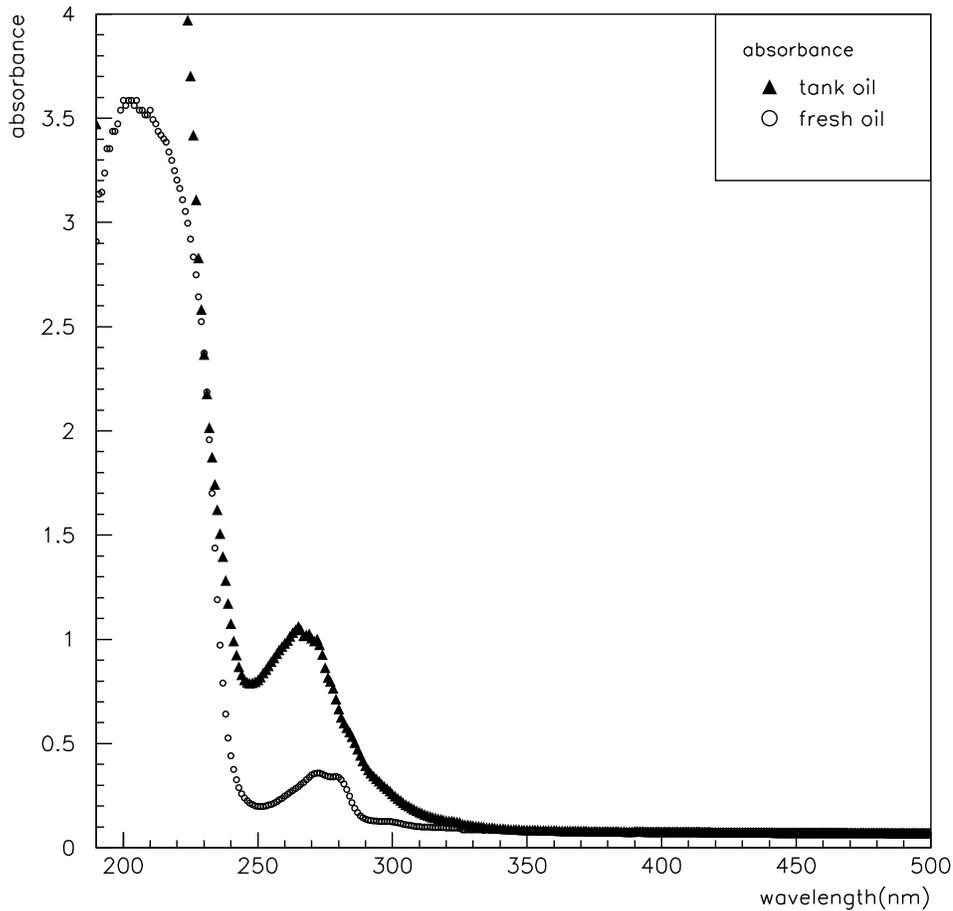} }
\caption{Absorbance spectra of fresh mineral oil from the Witco Corporation
(circles) and oil from the LSND storage tank (triangles).}
\label{fig:absorb}
\end{figure}

\begin{figure}[t,h,p]
\centerline{\includegraphics[width=5in,bb=23 160 532 655,clip]{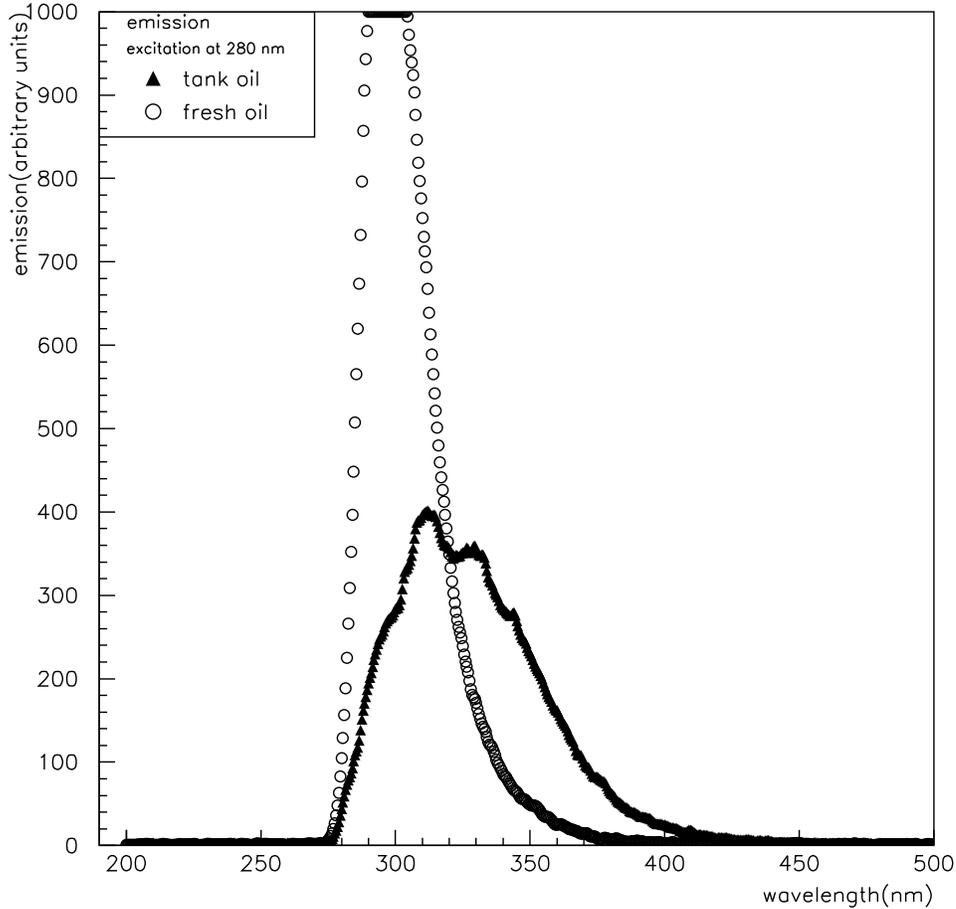} }
\caption{Emission spectra of fresh mineral oil from the Witco Corporation
(circles) and oil from the LSND storage tank (triangles).  The samples have 
been excited at 280~nm.}
\label{fig:emit}
\end{figure}

As a test of our procedures, oil
drawn from the LSND storage tank in 1998 was compared to newly-acquired ``fresh''
samples of the same Witco mineral oil. There is evidence that the black paint 
used in LSND caused some problems; they experienced a rising level of 
scintillation over their approximately six years of operation. The LSND storage
tank was prepared internally in the same way as its detector tank, but the oil
stored there for 6 years had no scintillator added. While fresh mineral oil is
odorless, the LSND storage tank oil had the smell of paint (which was also true
of the oil from the LSND detector when it was decommissioned). 
Figure~\ref{fig:absorb} shows the absorbance in 1~cm of oil. ``Absorbance'' is
defined as $-\log_{10}T$, where $T$ is the transmission. Note that an absorbance
of 1.0 in a 1~cm sample corresponds to an extinction length of only 4.3~mm. While
we are interested in maintaining extinction lengths of many meters at wavelengths
near the peak of the PMT efficiency around 400~nm, changes in the location and
structure of the turn-on of strong extinction proved to be a sensitive indicator
of contamination. Such an effect is clearly seen in Fig.~\ref{fig:absorb}.
Figure~\ref{fig:emit} shows emission spectra produced from oil samples when
excited by 280~nm light. The LSND storage tank oil has a significantly broader
excitation spectrum. Both of these measurements indicate the presence of
contaminants in the storage tank oil.

\begin{figure}[t,h,p]
\centerline{\includegraphics[width=5in,bb=0 275 565 565,clip]{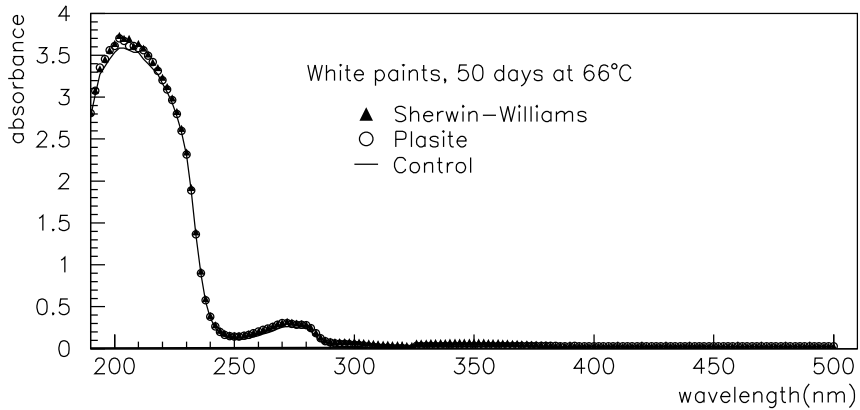} }
\caption{Absorbance spectra for oil in which metal plates painted with the
two white coatings used in the MiniBooNE tank were soaked for 50 days at $66^\circ$C. 
The solid line is the spectrum for the control sample of  oil.}
\label{fig:paint_oil}
\end{figure}

\begin{figure}[t,h,p]
\centerline{\includegraphics[width=5in]{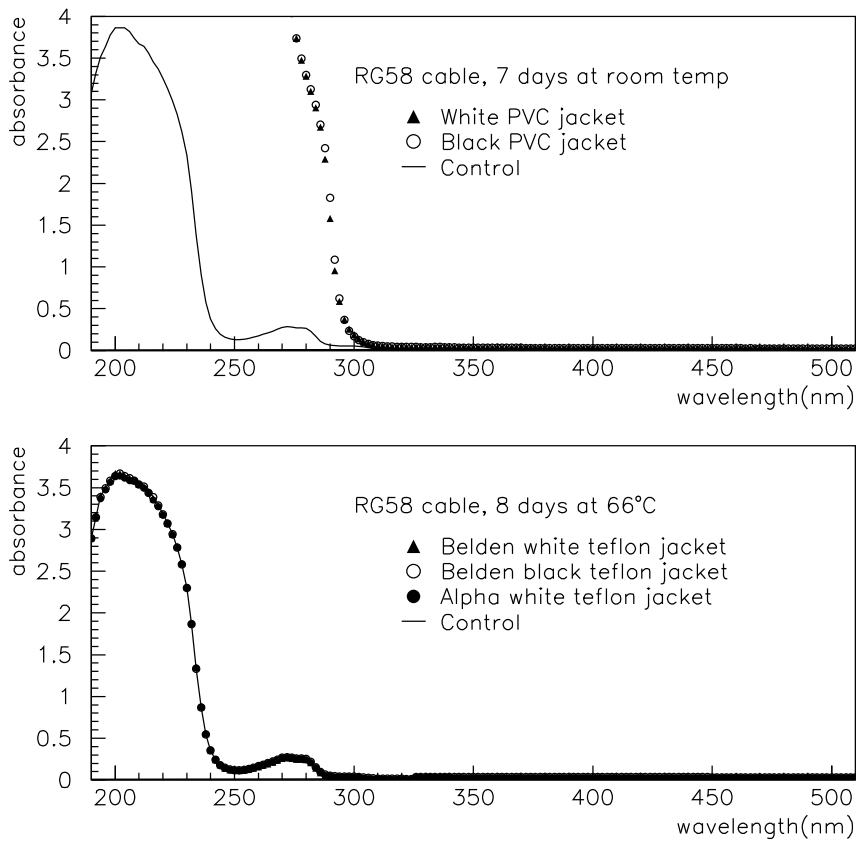} }
\caption{Absorbance spectra for oil in which new samples of RG58 cable were
soaked.  Top: black and white PVC-jacketed cables soaked for 7 days at room temperature.
Bottom: Black and white teflon-jacketed cables soaked for at least 8 days at $66^\circ$C.
(The spectrum shown for the Belden white cable, the type ultimately used in MiniBooNE, was
actually taken after an additional 69 days at varying temperatures.) 
The solid line is the spectrum for the control sample of  oil. }
\label{fig:rg58}
\end{figure}

\begin{figure}[t,h,p]
\centerline{\includegraphics[width=5in,bb=0 275 565 565,clip]{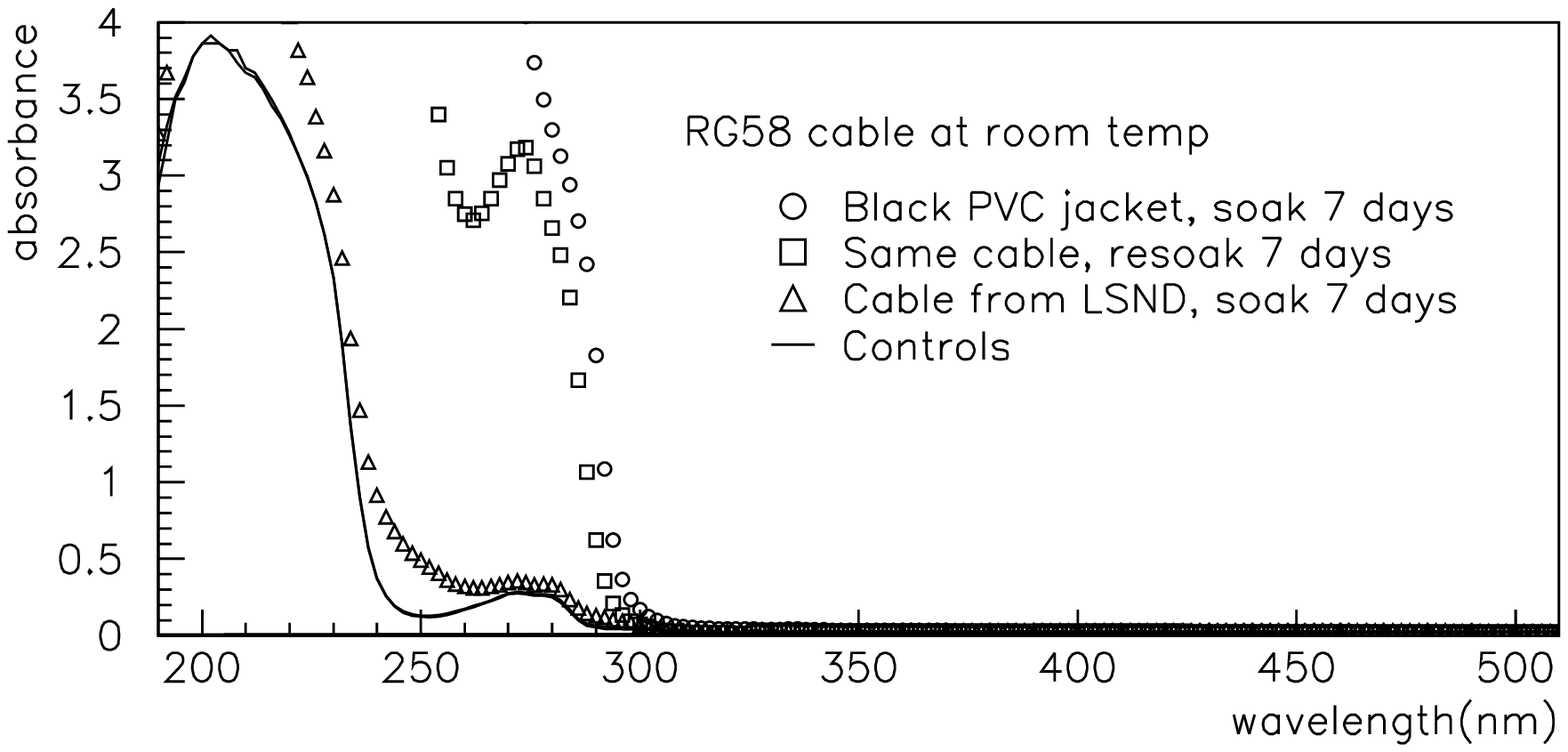} }
\caption{Absorbance spectra for oil in which black-PVC-jacketed RG58 cable were soaked.
The figure compares three samples: a new cable, soaked for 7 days 
(same data as the top panel of Fig.~\ref{fig:rg58}), the same cable resoaked
in fresh oil for 7 days, and a sample of cable from LSND (i.e. a piece of cable that was
immersed in the LSND tank during the duration of the LSND experiment), cleaned and soaked in
fresh oil for 7 days.
The solid lines are the spectra for the control samples of oil that accompanied the 
test samples.}
\label{fig:resoak}
\end{figure}

The surface coatings (the white of the veto volume and the black of the main
volume) present the largest surface area of material in the oil and therefore
present the greatest threat of contamination. 
The tank and PMT support structure coatings for MiniBooNE discussed in
Sec.~\ref{sec:paint} were chosen from those found not to change the optical
properties of the oil. 
Figure~\ref{fig:paint_oil} shows the absorbance results for the two white paints used
in the tank.

The PMT HV/signal cables also present a large surface area
to the oil. We found that PVC-jacketed cables, including the RG-58C/U with a
``non-con\-tam\-i\-nat\-ing'' jacket used in LSND, contaminate mineral oil. An example of
these results is given in the top panel of Fig.~\ref{fig:rg58}, 
which shows absorbance for oil in
which  new samples of PVC-jacketed RG58 were soaked for one week at room
temperature. Teflon-jacketed cables were found to be non-contaminating and Belden
88240 white (also called ``snow beige'') cables of this type were chosen for MiniBooNE. 
The bottom of
Fig.~\ref{fig:rg58} shows absorbance for oils in which teflon-jacketed cables
soaked at $66^\circ$C. One of the samples shown was allowed to soak for over two
months, with no contamination observed.

Further tests indicated that as a PVC-jacketed cable soaks, its contamination
rate decreases. Figure~\ref{fig:resoak} shows absorbance when the same
cable shown in Fig.~\ref{fig:rg58} (top) is soaked again for one week in fresh oil. 
Also shown in
Fig.~\ref{fig:resoak} is the absorbance for oil in which a cleaned sample of cable from
LSND ({\it i.e.,} the cable was immersed in the LSND tank during the life of the
experiment) soaked for one week at room temperature. The smaller impact on the
oil demonstrated in this figure permitted the reuse of the short lengths of
PVC-jacketed cable already attached to the bases of the PMT's recycled from LSND.

The bases and necks of the LSND phototubes were coated in Hysol potting
compound. New Hysol samples and samples taken from an LSND phototube were found
to contaminate oil. Therefore a new encapsulant epoxy was chosen; EP21LV from 
Master Bond (Hackensack, New Jersey). This epoxy meets FDA requirements for food
compatibility and was found to be non-contaminating in our tests. Both
MiniBooNE's old and new phototubes were potted in Master Bond epoxy; in the
case of the old tubes, the Master Bond was applied over the Hysol.

Hundreds of samples were tested to check other components. In addition, it was
confirmed that metals do not cause contamination of the oil at room temperature
and that they seem to catalyze oxidation at higher temperature.  Tests with 
aluminum and copper in oil at $66^\circ$C indicate contamination (the copper is
worse), and the contamination at $66^\circ$C is much smaller when the oil
samples are kept in a nitrogen environment. Therefore unpainted metal was
acceptable in MiniBooNE (we have some unpainted aluminum, but no copper), where
the oil remains cool and under nitrogen.

\section{PMT Support Structure}
The photomultiplier tube support structure (PSS) includes the hardware for 
supporting 1280 main and 240 veto photomultiplier tubes (PMT's), 
the optical barrier separating the main and veto oil volumes, 
the fixtures for support and strain-relief of the PMT cables,
and support for the various in-tank calibration and monitoring systems.
The dimensions of the sections of the PSS were chosen to uniformly distribute
the main PMT's over the inner surface of the optical barrier and the
veto PMT's over the tank wall.

It is not possible to distribute the phototubes over a sphere with
exact uniformity.  For reasons of structure and ease of installation,
the PMT's were deployed in evenly-spaced horizontal rows.  The number of
tubes in each row had to be even (since two tubes were placed on each panel of the optical barrier)
and was chosen to maintain horizontal spacing
that is as close to
uniform and as close to the vertical spacing as possible.  
The odd-numbered rows started with the first tube shifted by half 
the horizontal spacing of the tubes in that row. 
Fig.~\ref{fig:pmt_map} shows the resulting layout.

\subsection{Structure}
\label{sec:pss_struct}

\begin{figure}
\centerline{\includegraphics[width=5.0in]{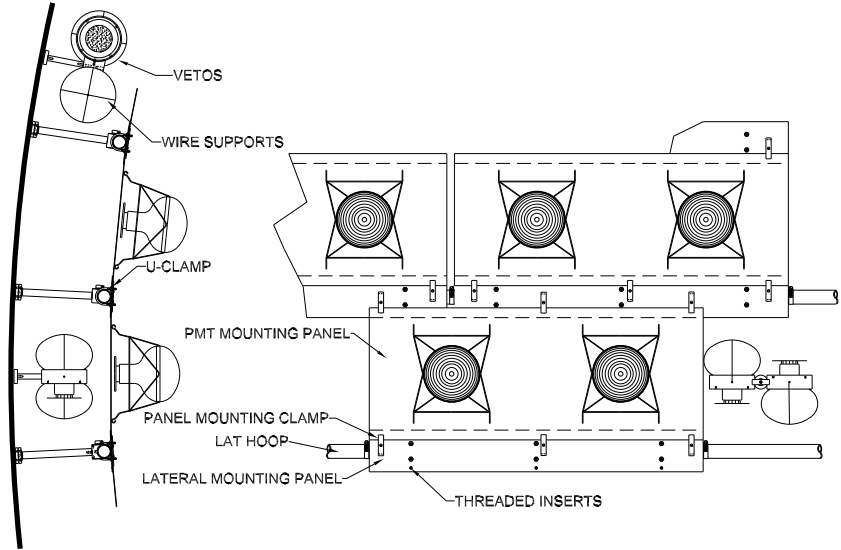}}
\caption{Overview of the Phototube Support System.}
\label{fig:PSS}
\end{figure}

   The basic design of the support structure is shown in Fig.~\ref{fig:PSS}.
The optical barrier panels are mounted on a set of 
latitudinal hoops each made of sections
of 5.08~cm (2~inch) diameter, 3.2~mm (${1\over 8}$~inch) wall, aluminum tubing.  
The sections are independently supported, but are
connected by sleeves to aid alignment.  Each hoop section is clamped
to 2.54~cm (1~inch) diameter steel struts, 
which are in turn bolted to 
bosses welded to the tank wall.  
The PMT's viewing the main oil volume are mounted on
the panels of the optical barrier.

Besides the obvious requirement of adequate strength, the PSS had to be
straightforward to install and tolerant of deviations of the
tank wall from a perfect sphere.  (Industry standards limit such
deviations to 1\%, or about 12~cm.)

A finite-element analysis of the tank showed that, while the displacements of
the tank wall when the oil was added would be, as expected, very small,
differential rotation of nearby struts could magnify this effect.  The largest
relative displacements, about 3~mm between the free ends of adjacent struts, 
occurred between strut pairs that straddled the region where a tank leg joined
the sphere.   Because the structure's ability to handle differential horizontal
motion was limited, we modified the boss layout to have bosses centered
horizontally on each leg,  reducing the maximum relative horizontal
displacement to less than 0.5~mm.

After the tank was complete and the bosses welded on, the Fermilab Alignment 
Group surveyed the boss locations. The survey indicated radial excursions  on
the order of 2~cm in the tank wall. We fabricated the struts in 0.64~cm
($\frac{1}{4}$~inch) increments in length, with  each boss assigned a strut
length based on the survey. The base of each strut is a disk with a recessed
ball-bearing at the center. This provides a pivot point, and the relative
tightening of the three mounting bolts could move the other end of the strut by
about an inch in any direction. This provided further compensation for
irregularities in the tank wall and  allowed generous tolerances in the
placement of the bosses on the tank wall, minimizing the cost of their
installation.

The panels of the optical barrier were made from 1.6~mm ($1\over{16}$~inch) 
aluminum sheet.  Each panel is approximately 1.2~m  wide by 0.6~m high and
holds two PMT's.   Experience with a full-scale prototype of a section of the 
PSS indicated that  flat panels easily conformed to the supporting hoops, so
that rolling them to the correct curvature was not necessary. When mounted,
each row of panels formed a section of a cone, with the whole optical barrier
approximating a sphere. (The sagitta formed by comparing the cones to the
sphere is 0.7~cm.) The panels were not mounted directly to the hoops. Instead,
they mounted to  15~cm wide strips which in turn mounted to the hoops using
U-bolts. The panels were attached to the strips by clips.   The lower clips
clamped the panels to the strips, while the upper clips captured the panels
without clamping them. Gaps of about 2~cm were left between adjacent strips and
panels in the same  rows. These gaps were blocked by narrow aluminum strips,
pop-riveted to the strip or panel on one side. This arrangement has a number of
virtues: 

\begin{description}
\item[$\bullet$] The overlap provides a robust optical barrier.  
With the detector complete, measurements with the
central laser flask showed the leakage of light to be $<10^{-3}$.
\item[$\bullet$] Variable gaps between panels horizontally and vertically
allowed loose mechanical tolerances, easing installation.
\item[$\bullet$] The use of the strips and clips decoupled the horizontal 
positions of panels in adjacent rows.  This and the gaps meant that 
installation variations did not accumulate as rows were added.
\item[$\bullet$] The fact that the panels were clamped only along one edge
(and, similarly, the inserts connecting the latitudinal hoop sections were only fastened
on one end) allowed the structure to shift slightly after installation, as could happen
when the 800~tonnes of oil was added or from thermal expansion.
\end{description}

To ensure uniform distribution of the oil during filling and circulation, 
there are
several ports through the optical barrier.  These are equipped with 
baffles to prevent light from crossing the barrier.

\begin{figure}
\centerline{\includegraphics[width=4.0in]{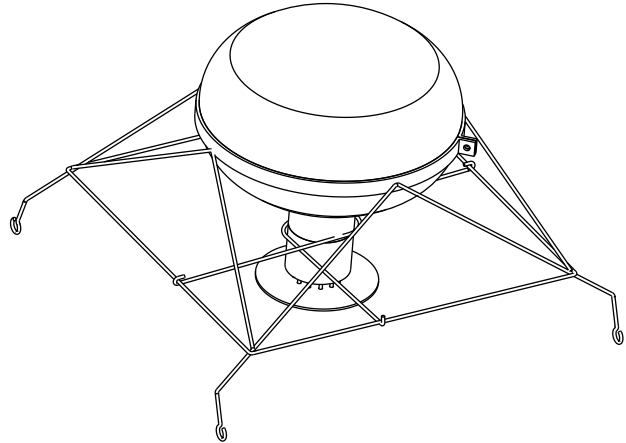}}
\caption{8-inch photomultiplier tube in LSND stand.}
\label{fig:LSND_stand}
\end{figure}

   Two PMT's were mounted to each panel using the existing PMT mounts from LSND
(see Fig.~\ref{fig:LSND_stand}), plus some new ones made to the same 
specifications.   The stands are made of 2.54~mm (0.1~inch) stainless steel
wire and an equatorial band of stainless steel sheet 1.27~cm (0.5~inch) wide
and 1.6~mm (0.065~inch) thick. The new PMT's differ from the old ones in the
diameter of the neck, so new centering clips were produced to accommodate
them.   The stands attached to blocks mounted on the panels.

\begin{figure}
\centerline{\includegraphics[width=4.0in]{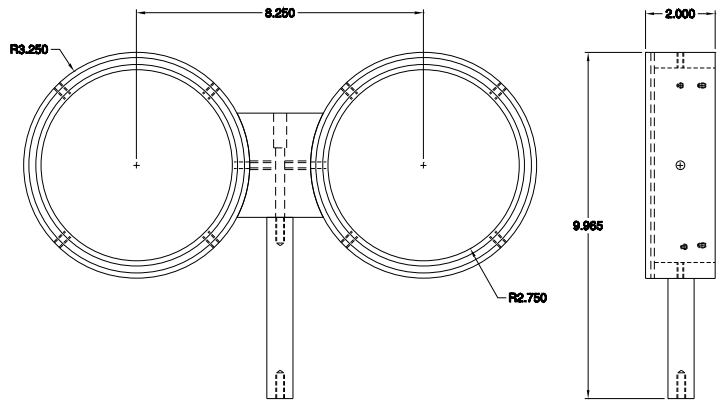}}
\caption{Veto PMT cluster.  The strut mounts to a boss welded to 
the tank wall.  The PMT's sit in the rings. Dimensions are in inches.}
\label{fig:veto_cluster}
\end{figure}

   The veto PMT's were mounted in opposite-facing pairs on struts attached to bosses welded to
the tank wall. The mounting scheme is shown in
Fig.~\ref{fig:veto_cluster}. Each phototube rests on a viton o-ring which in
turn sits on a step in the end of an aluminum pipe.  A cross of stainless steel
wire captures the globe of the tube against the o-ring, while nylon screws
center the neck of the tube in the pipe. The orientation of each cluster on its
strut could be varied to avoid  obstructions.  Monte Carlo studies showed no
differences in average light collection among various possible patterns in the
orientation.

\begin{figure}
\centerline{\qquad\qquad\qquad\includegraphics[width=6.0in]{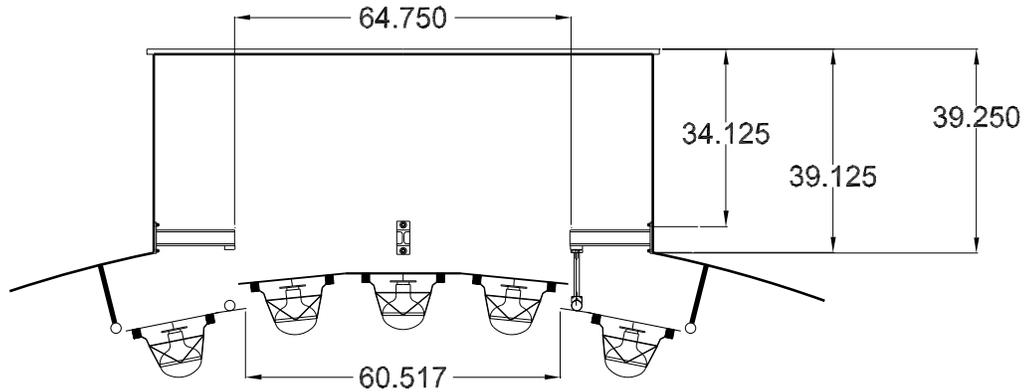}}
\caption{The top polar cap.  Dimensions shown in inches.}
\label{fig:north_cap}
\end{figure}

   To facilitate installation, 
the top and bottom ``polar caps'' were treated
specially.
Each holds the polar PMT and the next row of 6 PMT's.
The top cap also holds the scintillator cubes of the calibration system
(see Sec.~\ref{sec:cubes}).
Fig.~\ref{fig:north_cap} shows how the top polar cap was mounted.
For uniform light collection in this region, a white aluminum panel was mounted
on the I-beams shown in the figure, just under the surface of the oil in the
top access portal.  The
top veto PMT cluster was also mounted from these I-beams.

The main PMT's are about 55~cm apart; the veto
tubes are about 2~m apart.
A survey taken with the tubes in place showed the average radial location of
the apex (innermost point) of the main-tank tubes to be 548.2~cm, consistent
with the dimensions of the mounting hardware. 

\begin{table}
\centerline{\begin{tabular}{|c|c|} \hline
Part & Total weight (metric tons) \\ \hline\hline
panels/strips & 2.1 \\ \hline
lat hoops  & 0.9  \\ \hline
bosses/struts & 1.0 \\ \hline
hardware   & 0.05 \\ \hline
PMT's/bases & 1.5\\ \hline
LSND stands& 0.3\\ \hline
cables (in tank) & 0.7\\ \hline\hline
Total      & 6.6 \\ \hline
\end{tabular}}
\caption{Total weight, in metric tons, of various PSS components.}
\label{tab:PSS_weight}
\end{table}

Table~\ref{tab:PSS_weight} shows the weight of various PSS components.  These
can be compared to the approximately 37 metric tons of the tank shell itself, 
the approximately  800 metric tons of oil, and the total buoyant force of the
PMT's in oil of 4.6 metric tons.

\subsection{Surface finishes}
\label{sec:paint}

Surfaces of the detector were painted to provide high
albedo in the veto volume and low albedo in the main volume.
Reflection of light in the main volume 
of the detector can cause Cherenkov light to appear isotropic and delayed,
degrading particle identification. We thus wanted surfaces in the 
main volume to be non-reflective. 
In the veto volume, we simply 
wanted to maximize the total light collected by the sparse array of 
PMT's.
The inner surface of the tank wall, the bosses, the struts, the 
latitudinal hoops, and the outer side of the panels, strips, and overlaps 
of the optical 
barrier were painted white. The inner side of the optical barrier
was painted black.  Small parts on the inner surface of the optical barrier, 
such as the clips and PMT mounting blocks, were black-anodized.

\begin{figure}[h,t,p]
\centerline{\includegraphics[width=5in]{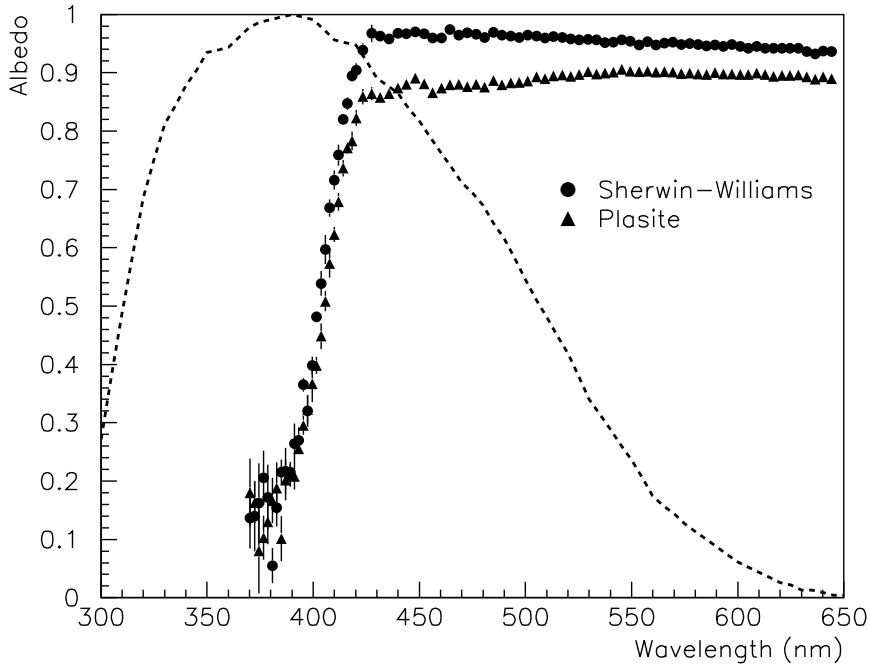}}
\caption{Albedo measurements for the white coatings used in the tank.
For reference, the dashed curve shows the PMT quantum efficiency, normalized to its maximum
value \cite{hamamatsu}.}
\label{fig:albedo}
\end{figure}

We measured the albedos of various paints in air using a tungsten lamp
and integrating sphere.    
All of the measured white coatings provided  
better than 80\% albedo at wavelengths above 425~nm, and
Fig.~\ref{fig:albedo} presents the 
measurements for the chosen white paints.
We also measured albedos in oil by immersing a sample,
a light source (an alpha emitter embedded in scintillator),
and a detecting PMT.  These in-oil measurements confirmed our
rankings of candidate surface treatments.

The steel tank was painted after construction with Plasite 9060, a white
epoxy coating.
Plasite Protective Coatings (Green Bay, Wisconsin) 
markets this tank lining 
to the food and beverage industries, saying that it 
will not impart taste or odor, and that it meets FDA requirements 
for direct food contact.  Our own tests indicated that it does not
contaminate mineral oil (see Sec.~\ref{sec:oil_compat}). 

The components of the phototube support structure were painted 
with coatings manufactured by Sherwin-Williams (S-W).  We found that these
coatings were non-contaminating.  In albedo tests of samples gathered 
from various manufacturers, the S-W white offered the
highest albedos, whereas the S-W black offered the lowest.  
The white is an aircraft paint (F91W26/V93V28/V93V2 Jet Glo Gloss Polyurethane White) 
and the flat 
black is a military paint (F93B102 Flat Black Moisture Cure Polyurethane).
Both were applied over a primer (E72AC500/V66V503 Hydralon P Waterbased Epoxy Primer (Grey)).

Some of the smaller pieces of the PSS (panel clips, the blocks to 
which the PMT stands attach, calibration cubes) were
black anodized.  This process was shown to be non-contaminating, and 
is a fast, cost-effective way to finish a large number of small pieces. 
However, a smooth black anodized aluminum surface offers fairly 
high albedo at large incidence angles; to achieve lower albedo the
surface must be scuffed (e.g., sandblasted) prior to anodizing. 
Anodizing was considered for the panels and rejected because of the
need for low albedo over the full range of incidence angles and the
concern that sandblasting might curl the panels.  
The smaller pieces were deburred in a tumbler, which 
left them slightly scuffed, making them ideal candidates
for anodizing. 

\subsection{Installation}
\label{sec:PSS_install}

   The installation of the PSS began in January 2001 and was completed in October 2001.
Installation was done from scaffolding 
supplied and installed by Bartlett Services, Inc., consisting 
of six levels with an internal stairwell.  

With the scaffolding in the tank, the cable bundles were installed in their flanges.
PSS/PMT installation then began on the top scaffold level with the PMT row below the 
top polar cap.  Each scaffold level was
lowered or removed as the PSS and PMT's above that level were installed and tested
to allow access further down the tank wall.  
The first steps at each level involved installing the latitudinal hoops to the
desired height using a rotating-laser level that traced a horizontal
line around the tank.
The height reference was a steel measuring tape hanging from the top access
portal.  It was calibrated by using the laser level to transfer the surveyed
coordinates of many bosses to the tape.

At regular intervals, the tank ports were closed and
the installed PMT's and
cabling were tested at high voltage using a 
temporarily-installed LED flasher system.

When nearly all the PMT's were installed, the Fermilab Alignment Group surveyed the 
positions of about 40\% of the main tank tubes.  
The survey showed that the horizontal rows had an rms deviation from the desired
polar angle of 0.08 degrees, corresponding to 0.7~cm at the radial location of the tubes.
The rms polar angle of the individual tubes within a row was of the same order.
Horizontally, the rows were oriented to within typically 0.25 degrees of the desired 
azimuthal angle, corresponding to about 2~cm.  
The radial position of the tubes had an rms of 0.9~cm.

\section{Utilities}
\subsection{Cable Plant}
\label{sec:tank_cable}

The tank cable run links the photomultiplier tubes to the preamplifiers, located
in crates in the tank access area next to the top tank access port. A single
RG-58 cable both supplies high voltage to a PMT and carries the signal from PMT
to preamp. As described in Sec.~\ref{sec:oil_compat},  teflon-jacketed cable,
Belden 88240, is used for compatibility with the mineral oil. This cable has the
further advantage of being plenum-rated. It is thus suited for running under the
computer floor to the preamps. As most of the run is in the veto section of the
tank, cable with a white jacket is used. The cable terminates at the preamp end
in a SHV jack. 

The penetration of the tank wall is through airtight feedthroughs bolted to the
flange on the tank access portal. The oil level in the access portal remains
below these flanges; the seal is to maintain the nitrogen atmosphere in the
tank. Each flange accepts eight bundles of 49 cables.

To maintain uniformity in pulse shape from tube to tube, cables of a fixed 
length, 30.5~m (100~ft), were used for all channels, despite the disparity in
the length of the cable runs to the top and bottom of the tank. Each cable is
captured in the feedthrough with only the length necessary to reach the PMT
extending from the inner side.

Each cable bundle serves the main and veto PMT's in a vertical slice of the tank
from top to bottom. Cables for the main PMT's cross the veto region and
penetrate the optical barrier through pairs of holes near the center of each
panel. After passing through the optical barrier, the cable was spliced to the
1.5~m length of black-jacketed cable that is permanently attached to the PMT
base. The splice was made with a coaxial splice kit (Raychem B-202-81, formerly
D-150-0071). This system was used in LSND and found to be very reliable. Of the
13 PMT failures that occurred in the first five years of MiniBooNE running, two or fewer
were attributable to splice failures. To seal out oil, the splice was covered
with a length of 9.5~mm (${3\over8}$~inch) teflon shrinktube (SPC Technology
SST-024), which shrunk onto viton o-rings on either side of the splice.

As each splice was made, it was tested. Prior to splicing, a pulser was
connected to the cable at the preamp end and the reflection (at about 370~ns)
was observed on an oscilloscope. Since the bases are back-terminated, the
disappearance of the reflection after splicing indicated both that the
splice was good and that the labels at the two ends of the cable matched.

Upon exiting the tank flange the cables were then routed under the raised floor
system surrounding the access portal to the preamplifiers. Prior to
preamplification, the PMT signals from single photoelectrons are small. The
raised floor between the tank access portal and the preamplifiers was thus
lined with copper mesh forming a Faraday cage to provide additional
shielding from noise pickup.

\subsection{Oil Plumbing}

\begin{figure}
\centering
\includegraphics[width=4.in]{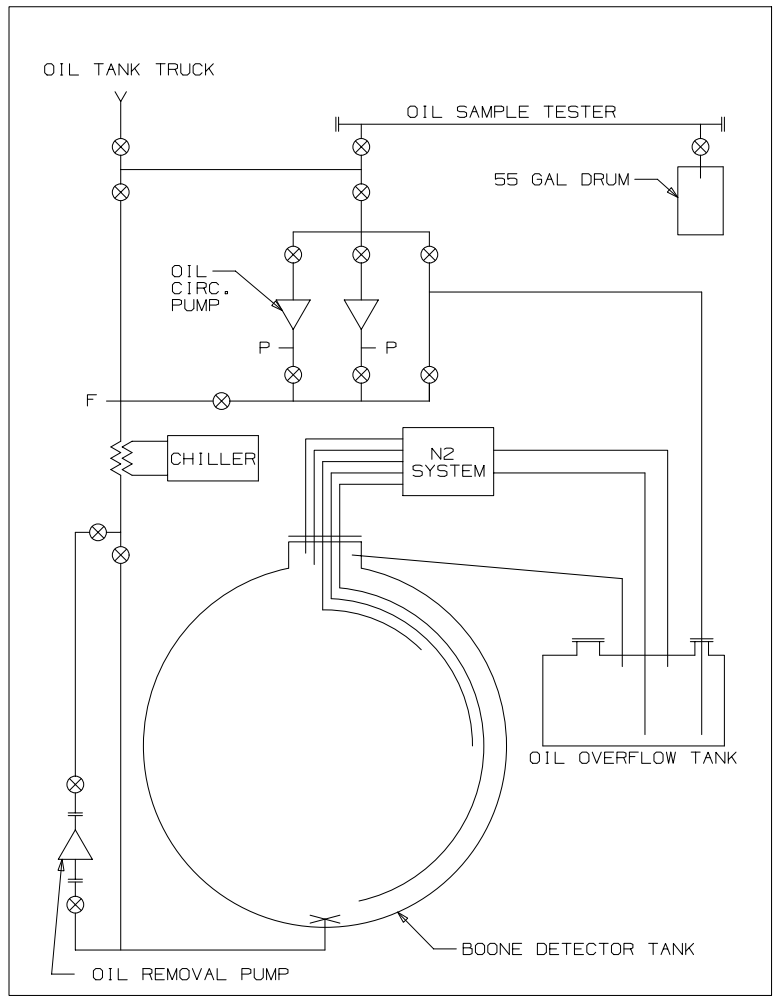}
\caption{A schematic figure of the MiniBooNE detector oil plumbing and
N$_2$ systems.}
\label{fig:oil_schematic}
\end{figure}

The MiniBooNE detector oil plumbing system was designed to enable filling,
recirculation, filtering, temperature control, and removal of the  the mineral
oil.  A schematic of the oil plumbing system  for the MiniBooNE
detector is shown together with the N$_2$ system in 
Fig.~\ref{fig:oil_schematic}. All components of the system are food-grade and all
oil pipes are stainless steel. Oil enters the detector tank  (capacity
950,000~liters) during fill or recirculation via a 7.62~cm (3~inch) pipe  that
attaches to the inflow penetration at  the bottom of the detector tank. During
decommissioning of the detector, this line will be used, together with a pump
located in the bottom of the detector vault, to remove the oil.

Flow studies indicated that, unless oil was  supplied directly to the main
volume of the tank, most of the circulation would  be through the outer (veto)
region. For this reason, oil enters the tank  through a coaxial, light-baffled,
fitting that delivers most of the oil to  the main volume. Another 7.62~cm
(3~inch) pipe is attached to the outflow flange on the detector tank access
portal.  This outflow line is routed to the oil overflow tank with a 1\% slope
so that any oil pushed out from the detector tank will passively flow to the
overflow tank. The outflow line is attached to the access  portal below the
level of the PMT cable ports so that the maximum height of oil is maintained at
a level below any other penetrations into the tank.

A 9,500-liter stainless steel oil overflow tank is located beneath the
detector  entrance hall and provides an overflow capacity of 1\% of the detector
tank volume.  The thermal expansion coefficient of the mineral oil was measured
to be $6.1\times10^{-4}~^{\circ}\mathrm{C}^{-1}$, therefore, the tank allows an
oil temperature change of $16^\circ$~C, well within any expected temperature
variation. 

During the filling stage, oil was pumped from the stainless steel tanker truck
into the bottom of the detector via the fill line. A stainless steel filter
installed in the inlet line removed most of the particulate matter as oil was
pumped in during delivery and during the later recirculation phase. The detector
main tank was completely filled and the overflow tank was filled to a third of
its capacity.

The oil inflow line passes through a heat exchange unit attached to a chiller
with 10~kW capacity. This arrangement allowed for cooling the oil during the
initial oil fill as well as during standard running in order to help remove the
approximately 600~W of heat generated by the PMT's. However, the temperature of
the oil during fill and subsequently, during standard operation running, has not
necessitated the use of this chiller.

\subsection{Nitrogen System}

The amount of emitted scintillation light and the light attenuation in the
detector  oil can vary strongly depending on the amount of oxygen dissolved in
the mineral oil.   Therefore, to maintain a stable detector medium, a nitrogen
environment is maintained throughout the oil plumbing system. In addition, the
(dry) nitrogen keeps water out of the detector tanks and lines and reduces the
problem of oxidation on exposed metal surfaces.

The nitrogen system is shown schematically in Fig.~\ref{fig:oil_schematic} and 
consists of a 1000~liter dewar of liquid nitrogen that provides gaseous N$_2$
via boiloff at 17~psi. The gas is routed to 2 microbubble (Point Four MBD600)
diffusers located at the bottom of the overflow tank.  Nitrogen is also provided
to one open line at the top of the overflow tank and four open lines terminating
at various heights within the detector tank.  Return lines were installed at
the top of the detector and overflow tanks and vented outside of the detector
hall.

Upon delivery, gas analysis indicated that there were approximately 30
(atmospheric)  cubic meters of oxygen dissolved within the mineral oil.  In
order to reduce this,  30 (atmospheric) liters per minute of N$_2$ was routed to
the microbubble diffusers  in the overflow tank while the mineral oil was
recirculated through the overflow and  detector tanks at 80 liters per minute. 
This situation was maintained for  one month.  A total of about 1 million
atmospheric liters of N$_2$ was bubbled  through the detector oil in order to
reduce the oxygen concentration by a factor of 10.  

After this procedure, the vigorous N$_2$ flow to the microbubble diffusers was
stopped and the oil recirculation halted.   Since that time, oil has not been
recirculated through the system and the N$_2$ flow has been set to approximately
1 liter per minute through the open lines to maintain the N$_2$ environment in
the tank. 

\subsection{Detector Environmental Monitoring}

A detector environmental monitoring system was installed to measure various
quantities during the commissioning and run phase of the experiment. These
include: air temperatures and relative humidities in the detector enclosure
and within the electronics racks; temperature, level, and flow rate of the
detector mineral oil; and pressure and flow rate of the N$_2$ gas.  These
quantities are measured at multiple locations throughout the detector and
enclosure. This system employed various commercial probes, displayed the values
on panel gauges, digitized the data using a standalone ethernet data acquisition
device (wedDAQ/100), and archived the values.

The average air temperature is $21^\circ$~C in the upper detector hall and
$18^\circ$~C in the  lower detector vault.  These values are stable to within
$2^\circ$~C throughout the year due to a HVAC system and the large thermal mass
of the concrete structure and earth overburden. The relative humidity in these
areas varies between 30\% and 60\% throughout the year. The exhaust air
temperature in the electronics racks is monitored closely as a loss of cooling
can lead to electronics failures. These probes are used with an interlock system
to shut down the power to the readout electronics if the temperature rises above
$40^\circ$~C.    The temperature of the mineral oil is quite stable without
additional cooling and is monitored at three locations within the detector tank.
The average temperature of the detector oil is approximately  $18^\circ$~C and
varies by $1^\circ$~C from the top to the bottom of the detector tank and by
$2^\circ$~C over the year. 

The oil level in the detector and overflow tanks are monitored with capacitive
level probes (Kubold NRF series). In the detector tank, the probe is located to
detect only the top 1~m of oil level.  During the fill stage, the oil level was
determined by the pressure in the N$_2$ lines within the tank.  During the oil
recirculation stage, the oil level in the main tank was set by the outlet pipe. 
When oil recirculation was halted, the oil level dropped slightly and varied 
with temperature throughout the year by approximately 20~cm.  This does not
affect the operation of the detector as the outer veto volume is
approximately 60~cm below the maximum oil level.

\section{Electronics and Data Acquisition}
\label{sec:elec_and_daq}
\subsection{High Voltage and Preamplifier System}

The PMT bases implement either a nine (old PMT) or ten (new PMT) stage design
using a Hamamatsu recommended voltage taper along the dynode chain for maximal
gain and reasonable linearity in the 1 to 100 photoelectron range of
illumination.  To remain compatible with the preamplification and high voltage
(HV) distribution system, the total base resistance is kept close to 17~MOhms. 
The voltage divider chain is made with carbon film, 0.5~W resistors and ceramic
disk capacitors and is entirely passive.  The anode circuit is back terminated
at 50~Ohms and is balanced to compensate for distortion caused by the capacitive
coupling in the termination circuit.  The base is designed to operate with
positive high voltage applied to the anode.  The cathode is at ground potential,
and isolated from the support frame to avoid ground loops.

Each HV/Preamp board handles the high voltage distribution and preamplification
of 24 PMT's. High voltage is distributed to each of the PMT's via 50~Ohm, equal
length, Teflon jacketed, 3000~V rated cables which also carry the PMT signal
(see Sec.~\ref{sec:tank_cable}). In order to achieve a uniform gain, the
PMT's were sorted into groups of 24 according to their gain's dependence on
voltage. An optimal voltage is then applied separately to each group, and
trimmed for each channel with a fixed resistor chosen to give that channel its
desired operating voltage.

The PMT signals are AC decoupled with a 0.01~$\mu$F, 3000~V, ceramic, HV
capacitor and are then amplified by a factor of 20 with an AD9617 op-amp in
non-inverting external feedback mode. To achieve low noise, this amplification
is done in a Faraday cage and as close as possible to where the HV cables exit
the tank.  The amplified signals from the HV/preamp boards are carried by
equal length, 50~Ohm, coaxial cable to the rear of the 8-channel DAQ charge and time (QT)
boards.  The racks housing the HV/preamp boards are located at least 3~m away from
the data acquisition electronics so as to mitigate digital noise effects.

\subsection{Signal Processing}
\label{sec:QTmeasurement}

The electronics digitizing system was designed to collect charge from each PMT,
digitize the time and charge amplitudes, store these for a substantial period,
and read them out when an external or internal trigger is received.

MiniBooNE re-used the LSND PMT electronics~\cite{Overview::LSND_NIM:1997}. All
power supplies, crates, electronic cards, and cables were checked and
refurbished before installation at Fermilab.

The data acquisition electronics are housed in a total of fourteen VME
crates: ten 9U QT crates process the 1280 tank PMT's, two 9U QT crates
process the 240 veto PMT's, and the trigger is housed in a 6U crate. One 9U QT
crate is used for high resolution time inputs from the calibration
systems (see Sec.~\ref{sec:calib}) and beam monitoring devices.
All crates conform to the VME standard ``A24:D32 Standard Non-privileged
Data Access''.

The system is driven by a 10~MHz rack-mounted GPS-disciplined clock, that feeds
into the trigger crate and is distributed to the other VME crates via
equal-length twisted pair ECL cables.  Central to data processing is the fact
that all digital information is stored on local channel memory that is addressed
by the lowest 11 bits of the GPS clock.  This Time Stamp Address (TSA) is
distributed via the TSA bus that daisy chains all crates together, thereby
keeping all hit information synchronized.

\begin{figure} 
\centerline{\includegraphics[height=4.in]{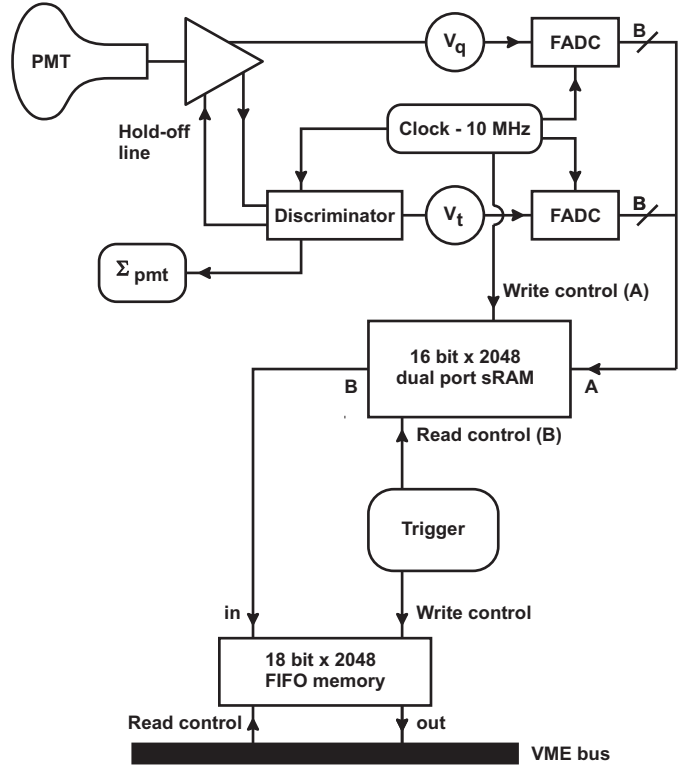}} 
\caption{A channel level block diagram of the front end QT electronics.}
\label{Overview::figure:MB_electronics_schematic}
\end{figure} 

The charge ($q$) and time ($t$) information from each PMT is processed on
eight-channel QT cards, of which there are sixteen in each crate. A channel
level block diagram of this system is shown in
Fig.~\ref{Overview::figure:MB_electronics_schematic}. If a PMT registers $q$ $>$
2~mV, a leading edge comparator discriminator fires immediately (asynchronous to
the clock) registering a hit and generating a 200~ns hold-off gate synchronized
to the clock. The PMT base resistor and gain are set such that 2~mV corresponds
to $\sim$0.1 photoelectrons.

\begin{figure}
\centerline{\includegraphics[height=4.in]{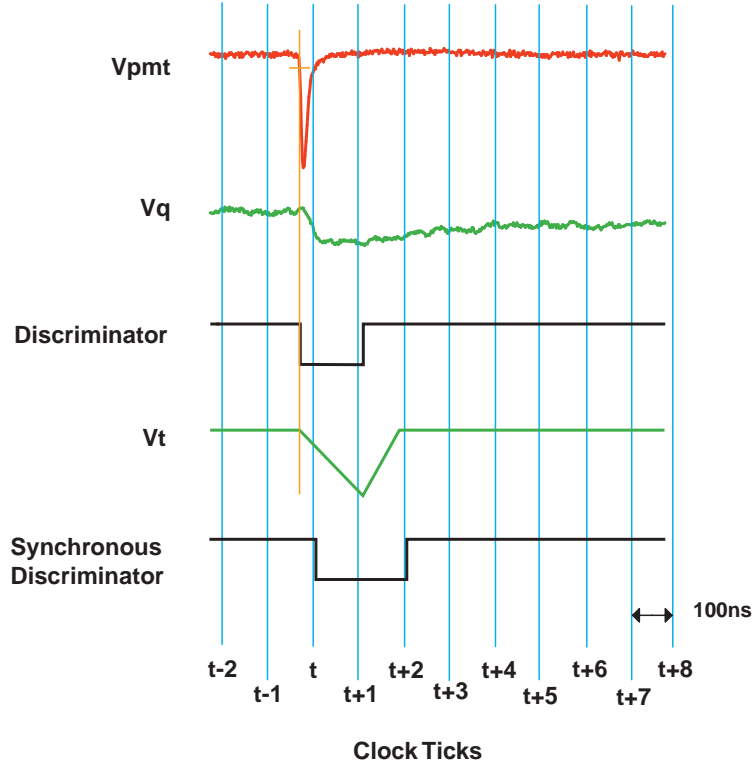}} 
\caption{MiniBooNE electronics traces and  gates for a single channel which show
how a PMT hit is processed by the electronics. The top trace ($V_{PMT}$) shows
the signal of a PMT hit crossing the 2~mV threshold, with the integral ($V_Q$)
of the pulse in the next trace.  $V_Q$ will bleed off over $\sim$1~$\mu$s. 
Next, the discriminator gate responds to $V_{PMT}$ crossing threshold, begins
the time ramp, $V_T$, and signals the synchronous discriminator to fire on the
next available clock pulse. ADC's digitize the integrated charge ($V_Q$) and
time ($V_T$) traces on every 100~ns clock tick.}
\label{Overview::figure:Quad}
\end{figure} 

Figure~\ref{Overview::figure:Quad} illustrates the signals produced by the QT
cards in response to a PMT signal. The PMT anode pulse is connected to an
integrator-stretcher circuit that convolves it with an exponential with a decay
constant of $\sim$700~ns. The digitized value of this stretched voltage at each
10~MHz clock tick is the $q$ measurement.  The empirical shapes of the
integrated charge bleed-off curves are extracted from MiniBooNE data, one for
LSND PMT's and one for new PMT's. The single photoelectron resolution is
approximately 4 bits, with a full scale of 20-30 photoelectrons.  The time
measurement is obtained by using a fixed transistor ramp circuit initiated by
the asynchronous discriminator firing that quickly returns to baseline after two
subsequent clock ticks. The ramp voltage, digitized at each 10~MHz clock tick,
gives the $t$ measurement. Two successive $t$ measurements determine the voltage
time slope. This is then extrapolated to the baseline voltage and gives the fine
time relative to the nearest 10~MHz clock tick. It has a least count of 0.8~ns.
The charge and time baseline distortions are less than 0.5\% during normal
running. The delay before the discriminator for a given PMT can fire again is
set by the synchronous discriminator and is between 200~ns and 300~ns depending
on the phase of the initial hit relative to the 10~MHz clock.

The digitization is done synchronously with the 10~MHz clock by 8-bit flash
ADC's. The $q$ and $t$ ADC values are stored in a 204.8~$\mu$s (2048 entries
spaced 100~ns apart), circular, dual-port buffer indexed by the TSA of each
clock tick.  Dual-port refers to the ability of the buffer to process input and
output values simultaneously on each clock tick.  The ADC digitization operates
independently of the trigger, and the circular buffer allows just over
200~$\mu$s for the trigger to respond to the stored charge and time data before
it is overwritten.  During normal running, the data loss rate is less than 0.1\%.
The number of hit PMT's (fired discriminators) is summed on each clock tick for
each crate and broadcast to the central trigger where a detector sum is done
separately for both main tank and veto PMT's.

\subsection{Trigger System} 
\label{sec:trigger}

The MiniBooNE trigger logic is constructed in software using external
information from the accelerator and calibration systems together with detector
PMT multiplicity information. The software runs on a monoboard computer housed
in a VME crate that also contains the (custom-built) trigger electronics
modules~\cite{Overview::LSND_NIM:1997}.  The information used for triggering
decisions is obtained from these modules.

Two trigger sum cards collect main and veto PMT multiplicity information from
$\Sigma_{PMT}$ cards that reside in each QT crate. The $\Sigma_{PMT}$ cards
compute the digital sum of the number of hits on the crate and transmit that sum
on a flat ECL cable to one of the trigger sum cards, one for main tank PMT's and
one for veto PMT's. The trigger sum cards then form global PMT multiplicities and
pass that information to the trigger memory card where seven different,
jumper-set, comparator levels (five for main, two for veto) are tested. The
trigger memory card also accepts three external inputs for NIM signals which are
used to input accelerator and calibration trigger signals. These external
signals were ``multiplexed'' to allow multiple different trigger types  to be
transmitted on the same input.  This is possible by encoding the trigger type
information in the NIM pulse length. (See
Table~\ref{Overview::table:comparator_settings} for a complete list of trigger
inputs.)

\begin{table}
\centerline{\begin{tabular}{|l|l|l|} \hline
Input        & PMT Hits            & Purpose		       \\ \hline\hline
External 1   & n/a		   & Beam to MiniBooNE         \\
External 2   & n/a		   & Strobe, NuMI, Debuncher   \\
External 3   & n/a		   & Calibration event         \\ \hline
Comparator 1 & Tanks hits$\geq$10  & Activity Monitor	       \\
Comparator 2 & Tanks hits$\geq$24  & Michel electron	       \\
Comparator 3 & Tanks hits$\geq$200 & High-Energy Neutrino      \\
Comparator 4 & Tanks hits$\geq$100 & Neutrino candidate        \\
Comparator 5 & Tanks hits$\geq$60  & Supernova $\nu$ candidate \\
Comparator 6 & Veto  hits$\geq$6   & Cosmic Veto	       \\
Comparator 7 & Veto  hits$\geq$4   & Cosmic Activity	       \\ \hline     
\end{tabular}}
\caption{The MiniBooNE external trigger inputs and comparator settings.} 
\label{Overview::table:comparator_settings} 
\end{table}

When any of the external input bits are set or the comparator bits indicate some
activity in the detector (tank hits$\geq$10 or veto  hits$\geq$6), an entry
containing the states of all these bits is placed onto the trigger FIFO register
(2048 words deep), along with a 16-bit time-stamp. FIFO status bits notify the
software running on the trigger monoboard computer when the FIFO has data ready
to process. The rate of trigger FIFO loads is approximately 150~kHz. The
external input and comparator bits, along with their time history are used by
the trigger software to determine if a trigger should be set.

When the software determines that a trigger has occurred, a contiguous string of
TSA's, corresponding to the time window of interest, is built and written to the
trigger broadcast card which then transmits that sequence of TSA's to the QT
crates. The number of TSA's in the string varies according to the type of
trigger. Receiver cards in the QT crates distribute the TSA list to the QT cards
which initiates a transfer of data from dual-port memory into the QT FIFO's.  As
described in Sec.~\ref{sec:QTmeasurement}, the depth of the dual-port RAM allows
for 205~$\mu$s of data buffering while the triggering decision is made.  If that
time is exceeded, the event is flagged in the data stream. The trigger hardware
settings and software code was tuned to so that this loss rate was less than
0.1\% during normal running.   

The principal physics trigger for this experiment requires only that beam was
sent to the MiniBooNE target regardless of whether or not there was any activity
in the detector.  This information is obtained from the accelerator clock
signals via the Fermilab ACNET (Accelerator Network). This signal is routed to
the trigger memory card on a dedicated input which was not multiplexed. When a
beam trigger is set, a string of 192 contiguous TSA's (19.2~$\mu$s) is broadcast
to the QT cards. This time window was adjusted to begin approximately 5~$\mu$s
before the arrival of the 1.6~$\mu$s beam spill. In this way, activity in the
detector before and after the beam spill is recorded  with minimum bias. A beam
trigger initiates a beam holdoff window (in the trigger software) that inhibits
the acceptance of subsequent triggers for 20~$\mu$s.  

Other trigger types were formed in order to produce special purpose data sets.
These included triggers that were used to collect neutrino events from other
sources at Fermilab (NuMI and the debuncher), to search for Supernova neutrinos,
to provide calibration samples, and to monitor  detector performance.  These
trigger types are summarized in  Table~\ref{Overview::table:EventTypes}. The
total trigger rate is $\sim$26~Hz, of which up to 5~Hz is  due to beam triggers.
The remainder of this section describes the conditions required for and the
purpose of each trigger type.

\begin{table}
\centerline{\begin{tabular}{|l|r|r|r|} \hline
Trigger         & Rate (Hz)   & Prescale & Time Window ($\mu$s) \\ \hline \hline
Beam            &	 2-5  &        1 & 19.2   \\ \hline
Beam $\gamma$   &  $\sim$0.1  &        1 &  3.2   \\ \hline
Beam $\beta$    &  $\sim$0.3  &        1 &  3.2   \\ \hline
Strobe          &	 2.0  &        1 & 19.2   \\ \hline  
Strobe $\gamma$ &	 0.1  &        1 &  3.2   \\ \hline
Strobe $\beta$  &	 0.3  &        1 &  3.2   \\ \hline
NuMI            &	 0.5  &       1  & 19.2   \\ \hline
BigNu           &	 0.66 &       1  & 19.2   \\ \hline
Michel          &	 1.2  &      600 & 19.2   \\ \hline
Supernova       &	 9.9  &       1  &  3.2   \\ \hline
Tank            &	 0.4  &    90000 & 19.2   \\ \hline
Veto            &	 0.4  &     5000 & 19.2   \\ \hline 
Laser           &	 3.33 &        1 &  9.6   \\ \hline
Fake beam       &  $\sim$0.5  &        1 &  9.6   \\ \hline
Cube            &	 1.1  &        1 & 12.8   \\ \hline
Tracker         &	 0.7  &      170 & 12.8   \\ \hline
\end{tabular}}
\caption{MiniBooNE trigger types. The total trigger rate is $\sim$26~Hz.} 
\label{Overview::table:EventTypes}
\end{table}

\begin{itemize} \item {\bf Beam}: The primary physics trigger for the
experiment. Requires the ``beam to MiniBooNE'' signal from the accelerator clock
and does not depend on PMT activity in the detector. 

\item {\bf Strobe}: Generated by a pulser running at 2~Hz, this trigger
produces an unbiased sample of beam-off triggers for background studies.

\item {\bf NuMI}: Set via the accelerator clock signal for beam sent from the
Main Injector to the NuMI beamline. It is analogous to the Beam trigger and 
allows MiniBooNE to detect off-axis neutrinos from the NuMI
target~\cite{numi_alexis}. The Fermilab accelerator clock cycles are such that
this trigger does not overlap with MiniBooNE Beam triggers.

\item {\bf BigNu}: Requires tank hits~$\geq$200 and veto hits~$<$6 and occurs
outside the beam and laser time windows. This trigger is set for neutrino
candidate events that are not induced by the Booster or NuMI neutrino
beams. 

\item {\bf Michel}: Requires tank hits~$<$200 and veto hits~$<$6 and the
signature of a cosmic-ray muon (tank hits~$\geq$100 and veto hits~$\geq$6)
between 3 and 15~$\mu$s prior. The trigger
collects a sample of cosmic-ray muon-decay electrons (``Michels'') in the
detector. The total rate of muon-decay electrons in the detector was
approximately 2~kHz, the requirement on a previous muon signal and prescaling
reduced the rate of this trigger to 1.2Hz.

\item {\bf Supernova}: Requires tank hits~$\geq$60 and veto hits~$<$6 and that
the time to a previous event with tank hits~$\geq$100 or veto hits~$\geq$6 was
larger than 15~$\mu$s. This trigger was established to search for neutrinos from
supernovae \cite{supernova_beacomb}. The tank hit requirement sets a lower
energy threshold of ~$\sim$20~MeV, higher than optimal for a Supernova search,
but necessary due to the high rate ($\sim10$~kHz) of $\beta$-decays 
in a near-surface detector. A supernova within 10~Kpc of the Earth
is expected to produce a burst of these triggers at a rate at least 5 times
greater than background. The rate of this trigger is $\sim10$~Hz.

\item {\bf Beam Gamma}: Requires tank hits~$\geq$10, tank hits~$<$100, and veto
hits$<$4, within 1~ms of a beam event that contains a neutrino candidate,
and no prior tank or veto activity within 15~$\mu$s. This trigger was designed
to obtain a sample of $^1H(n,\gamma)D$ events (neutron capture) with the photon
at 2.2~MeV, where the capture lifetime is 186~$\mu$s. 

\item {\bf Beam Beta}: Requires tank hits~$\geq$24 and veto hits~$<$4 within
30~ms of a beam event that contains a neutrino candidate, and no prior tank or
veto activity within 15~$\mu$s. This trigger is designed to select events from
excited nuclei with longer lifetimes that will typically beta-decay within tens
of ms after the neutrino event.

\item {\bf Strobe Gamma}: Same conditions as the Beam Gamma trigger, but is
initiated by a Strobe trigger. Used to measure accidental rates. 

\item {\bf Strobe Beta}: Same conditions as the Beam Beta trigger, but is
initiated by a Strobe trigger. Used to measure accidental rates. 

\item {\bf Tank}: Requires tank hits~$\geq$60 and is prescaled. Used for
detector monitoring.

\item {\bf Veto}: Requires veto hits~$\geq$6 and is prescaled. Used for
detector monitoring.

\item {\bf Laser}: Requires a calibration laser signal (Sec.~\ref{sec:laser}).
Used for detector calibration. 

\item {\bf Fake Beam}: Requires a calibration beam signal. The laser is pulsed
when the Booster extracts beam to the Main Injector instead of the MiniBooNE
beam target. This provides a check on the effect that the accelerator operations
may have on behavior of the PMT's and electronics.

\item {\bf Cube}: Requires a calibration cube signal which is set when the
calibration cubes have detected a cosmic ray muon candidate. (see
Sec.~\ref{sec:cubes}).

\item {\bf Tracker}: Requires a calibration tracker signal which is set when the
cosmic-ray muon tracker has hits in all four tracking planes (see
Sec.~\ref{sec:cubes}).

\end{itemize}

\subsection{Data Readout and Processing}
\label{sec:raw_data}

Every VME crate in the data acquisition (DAQ) system, including the trigger,
contains a Motorola MVME 2604 single board computer, referred to as a
``monoboard''. This serves as the VME master as well as the resident 300~MHz
PowerPC CPU to read data from the VME bus, send VME commands, run programs, and
connect to the local DAQ network to send event data via 100~Mbit Ethernet.  The
boards are not equipped with hard drives, and must instead mount to the disks
on the central DAQ computer.  Each monoboard has a Linux kernel with only
essential services and 64~Mbytes of memory. Each also has a serial port for
the console and an Ethernet port used for data transfer and communications.
Three rack-mounted computers are used for central data acquisition, processing, and
monitoring. They also run the scripts associated with permanent data archival
to Fermilab's Feynman Computing Center. All fourteen of the monoboards and the
DAQ computers are connected via a 100~Mbit Ethernet hub to the internal
MiniBooNE DAQ network.

The major components of the DAQ software are the QT software, trigger software
(discussed in the previous section), the event-builder, run control,
diagnostics, and archiving.

The QT software is a suite of C programs which run on the monoboards.  The
software uses address references to access the different VME functions.  At the
lowest level, a kernel driver checks the status of the TSA receiver and QT
FIFO's 60 times/second. If the ``empty flag'' on the FIFO is not set, then the
driver unloads data from the FIFO (via a VME interface). The advantage of using
a driver instead of a conventional program is that it interrupts the running of
any other code to unload FIFO's which fill rapidly making this a high-priority
task.  The driver makes the data available to the higher-level software for
further processing.  Next, each of the charge, time-ramp, and associated TSA
data words are checked for inconsistencies. If any are found, error flags are
set. Otherwise the data is compressed into good PMT hits, packaged with a
header (TSA address) and sent via Ethernet to the central DAQ computer. 
The header contains the information needed to match the hits from all of the QT
monoboards sending data.

The data required to reconstruct the charge and time for a PMT hit consist of
the set of four consecutive digitized $q$ and $t$ ADC values starting with the
100~ns clock tick immediately preceding the hit. These eight values form a data
``quad''. Most PMT hits come in the form of quads, but occasionally, for
high-energy hits, the charge ADC will saturate and, depending on the amplitude,
will take as much as a few microseconds to stabilize. This condition is detected
in the packaging software and causes it to extend the hit data up to 400~ns
after the ADC values become unsaturated. PMT hits are detected in the online
software through the presence of a ramp in the $t$ ADC values. Data quads
without a time ramp are ignored.  This reduces the data stream by a factor of
100.

The first step in the procedure to convert a quad into PMT hit charges and times
uses the first three $t$ ADC values of the quad to determine $t_{raw}$, the raw
time (in ns) that the discriminator fired relative to the clock tick that
preceded it. Since a hit in the detector produces a known integrated charge
shape, $V_q(t)$, in the electronics, the newly determined raw time can be used
to predict the expected $q$ ADC values for a given charge deposition. Hence, the
raw charge, $q_{raw}$, is determined by finding the normalization of $V_q(t)$
that best fits the $q$ ADC values associated with that hit. A charge-dependent
time slewing correction, $t_{slew} (q)$, is taken from a lookup table indexed by
$q_{raw}$; separate tables are used for the old PMT's and the new PMT's. The
charge for saturated hits is determined by measuring the relaxation time of the
integrator, since the charge amplitude is correlated with the integration
relaxation time. This procedure is applied during offline data processing.  

The event-builder collects Ethernet data packets from the QT monoboards and the
trigger.  When it receives a hit or set of hits from a crate, a status array
indicates the existence of data from a crate.  The data come asynchronously
from the QT monoboards, and event-builder continues to listen for more data
until all monoboards have sent charge and time data with matching TSA
information.  The trigger software also sends a packet of data showing all
current trigger information that the event-builder merges with the data stream. 
It also performs integrity checks on the time information, and in particular,
compares the trigger packet with the QT data packet.  If a problem is found,
then a new run is started, including an automated reset of all running software
and VME devices. If not, the event is stored in binary format to disk. For
every 100 Mbytes of data, a new file is created and numbered to allow easier
manipulation and file transfer. At normal data taking rates, this occurs every
15 minutes, yielding a total data rate of $\sim$100~kBytes/sec.

Once a new data-file is available, analysis scripts run on the file to monitor
the health of the detector and subsystems. Thousands of plots such as PMT dark
rates, tank and veto hit sums, {\it etc.}, are made available to the shift personnel to
check the health of any aspect of the detector. Concurrently, the files are
streamed from their original binary format to ROOT files which are written to
tape along with the original files. The robotic tape system located at the
Fermilab Feynman Computing Center facility is used for all tape archives. 
MiniBooNE has recorded approximately 18 Tbytes of raw data in five years of data
taking.

Beam line, horn, and environmental data are merged offline with the detector
data. All of the data streams are coordinated by the run control software which
propagates time stamps and new run information to the other data stream's
acquisition software in order to maintain uniform run boundaries.  The most
important of these external data streams is the beam line data which are
acquired every beam spill using seven Internet Rack Monitors (IRM) and a
``multiwire'' front end.  Each IRM collects data for one system: beam positions
along the beam-line, beam loss monitors, beam position at target, target cooling
system, horn system, beam intensity, and magnet power supplies.  The
``multiwire'' front end records beam targeting information. Both the IRM's and
multiwire front end push the data to the beam line data acquisition system,
which bundles the data into a format allowing simple integration with the
detector data.  The state of the beam-line and proton beam can be reconstructed
for each event.  In particular, the number of protons incident on the target is
recorded for each spill which allows for a more precise understanding of the
neutrino flux.

\section{Calibration Systems}
\label{sec:calib}
\subsection{Laser Calibration System}
\label{sec:laser}

\begin{figure}
\centerline{\includegraphics[width=4.0in]{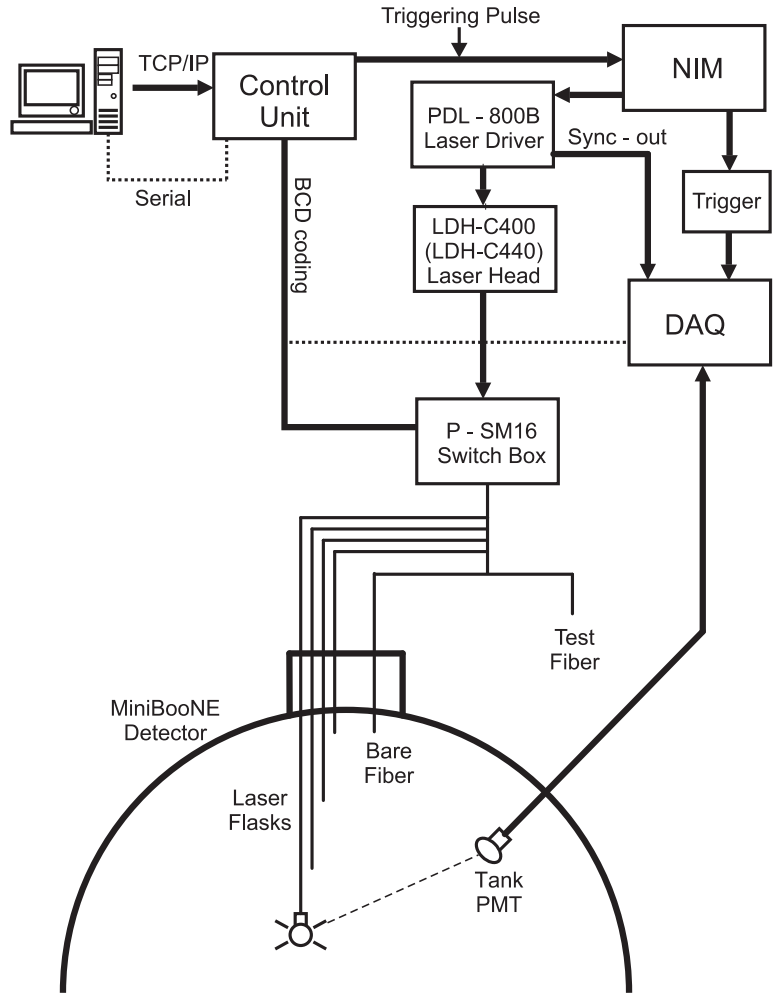}}
\caption{ Schematic diagram of the MiniBooNE laser calibration system.}
\label{fig:laser_diagram}
\end{figure}

The MiniBooNE laser calibration system consists of a pulsed diode laser (PicoQuant
GmBH PDL-800-B driver and LDH 375 laser head), and four dispersion flasks
installed at various locations in the detector. Short pulses ($<$1~ns) of laser
light with wavelengths peaked at 397~nm, are transmitted via optical fibers to
each of the dispersion flasks. The primary
purpose is to quantify and monitor individual PMT performance parameters. It
also allows for \emph{in-situ} monitoring of the oil attenuation length over the
lifetime of the experiment. A schematic diagram of the laser/flask system is
shown in Fig.~\ref{fig:laser_diagram}.  A switch box allows transmission of the
laser light pulses to one of the dispersion flasks via an optical fiber. Each
dispersion flask is 10~cm in diameter, filled with Ludox colloidal
silica~\cite{ludox}; laser light sent to a flask illuminates all of the PMT's
with approximately equal intensities. In addition to the four flasks, there is a bare
optical fiber that emits light in a cone of $\sim\mathrm{10}^{\circ}$ opening
angle, illuminating PMT's in a small circle near the bottom of the detector. It
is used to study light scattering in the detector. The locations of the four
flasks and the bare fiber are shown in Table~\ref{tab:laser_positions}; the
coordinate system's origin is at the center of the detector, the $z$-axis points
in the direction of the neutrino beam and the $y$-axis points in the vertical
direction.

\begin{table}
\centerline{\begin{tabular}{ |c|c|c|c|c| } \hline
Device     & $x$ (cm) & $y$ (cm) & $z$ (cm) \\\hline \hline
Flask 1    &   -0.3 &   -4.1 &    1.5 \\
Flask 2    &  144.9 &   96.1 & -126.4 \\
Flask 3    &    1.7 &   -0.8 &   83.7 \\
Flask 4    &  -80.0 &  203.9 &  -24.1 \\
Bare fiber &   82.0 &  540.0 &   65.0 \\ \hline
\end{tabular}}
\caption{Positions of laser light sources in the MiniBooNE detector.}
\label{tab:laser_positions}
\end{table}

During normal data-taking, laser light is sent to the central dispersion flask
(flask~1) at rate of 3.33~Hz, continuously and asynchronously with the
accelerator. The laser system is vetoed by a beam
extraction timing pulse and is thereby prevented from firing light into the
detector during a beam event. The laser intensity is set to produce a low PMT
occupancy; on average $\sim$3\% or 40 PMT's record hits in each laser event.

The raw PMT hit times described in Sec.~\ref{sec:raw_data} are transformed into
calibrated times using individual time offset constants, one for each
PMT. The raw PMT charges are normalized into units of photoelectrons by
individual multiplicative gain calibration constants, one for each PMT. These
calibration constants are extracted from the laser calibration data. The
time offset is used to account for transit time differences in each
PMT/QT readout channel.  These time offsets are calculated by measuring the
difference between the observed time of each PMT hit and the known time of the
laser pulse, with corrections for the laser light time-of-flight. The
gain calibration constant is used to account for individual
differences in the gains of PMT's.  The gain constants are calculated by
defining the mean of the $q_{raw}$ distributions from calibration laser events
to have the value of one photoelectron.  A correction for Poisson fluctuations
in the actual number of photoelectrons observed is made, using the measured
average occupancy for calibration laser events.  To account for charge dependent
time-slewing, a time slewing lookup table is created using calibration laser
events.  The table is indexed by the raw (not gain calibrated) charge,
$q_{raw}$, because it depends on the amount of charge seen by the ADC's, not on
the number of photoelectrons hitting the PMT.

Single photoelectron charge distributions are also extracted from the laser
calibration data (one for each type of PMT), and are used as inputs to the Mont
Carlo simulation of the detector.

\subsection{Cosmic Muon Calibration System}
\label{sec:cubes}

The muon calibration system hardware consists of a scintillator hodoscope
located above the detector and seven scintillator cubes deployed within the
detector signal region. A distinct signature is produced when a muon passes
through the muon tracker, stops in a scintillation cube, and decays. The muon
produces Cherenkov light that is detected by the tank PMT's as well as a
coincident signal in the scintillator cube. When it decays, the resulting Michel
electron also produces coincident signals as it exits the cube and traverses the
tank volume. Such events, where the location and momentum of the muon and
the origin of the electron can be independently determined from the muon
hodoscope and cube geometry, provide a means of tuning and verifying event
reconstruction algorithms.

The hodoscope has two layers of plastic scintillator, each with an $x$ and $y$
plane, that determine the positions and directions of muons entering the
detector.  The scintillator strip widths in the muon tracker and the separation
between the planes were chosen to give adequate coverage throughout the entire
detector and to have a reconstruction angular resolution that would be dominated
by the effects of multiple scattering of the muon as it passes through the
detector; the RMS resolution of the muon tracker itself is $\sim1.9^{\circ}$
(34~mrad).

\begin{table}
\centerline{\begin{tabular}{ |c|c|c|c|c|c| }\hline
Cube depth (cm) & $x$ (cm) & $y$ (cm) & $z$ (cm) &  $<Range>(\mathrm{g/cm^2}) $ &   $<T_{\mu}>(\mathrm{MeV})$  \\ \hline\hline
31.3            &  -60.8  &  540.7  &  15.1   &  28.  $\pm$ 1.               & 95.  $\pm$ 4.                \\
60.3            &   15.6  &  511.7  & -57.6   &  54.  $\pm$ 2.               & 155. $\pm$ 5.                \\
100.5           &   57.9  &  471.5  & -13.5   &  89.  $\pm$ 3.               & 229. $\pm$ 7.                \\
200.8           &  -18.6  &  371.2  &  59.2   &  174. $\pm$ 4.               & 407. $\pm$ 9.                \\
298.1           &   40.8  &  273.9  &  44.5   &  256. $\pm$ 4.               & 584. $\pm$ 9.                \\
401.9           &   40.8  &  170.1  &  44.5   &  344. $\pm$ 4.               & 771. $\pm$ 9.                \\ \hline
\end{tabular}}
\caption{Positions of cosmic muon calibration cubes.  The depth  is the distance
of the center of the cube below the optical barrier. $<Range>$ is the average
material, within the detector signal region, traversed by muons that pass
through the muon tracker and stop in the cube. $<T_{\mu}>$ is the kinetic energy
of the muon as it enters the signal region.  The range uncertainty shown is the
RMS of the range reconstructed for data events from the geometry of the
hodoscope and cubes. It is propagated into the kinetic energy uncertainty using
the range-energy table from Ref.~\cite{mu_range}, with an additional
contribution from straggling.}
\label{tab:cubes_positions}
\end{table}

The positions of the cubes in the detector vary from 15~cm to 400~cm from the
optical barrier, so that stopping muons from 20~MeV to 800~MeV can be measured.
Each 5-cm cube is housed in a sealed aluminum box
and has an optical fiber leading to a 1~inch PMT for readout.  The deepest cube
is larger, 7.6~cm on each side, to increase its signal rate.  One cube is
deployed with a special optical barrier that blocks light produced behind the
PMT's, reducing the total path length for muons stopping in it to $\sim$11~cm;
that cube is not used in the results discussed here. 
The positions of the cubes in the tank were surveyed by the Fermilab alignment
group; the reported error on their measurements is less than
3~cm~\cite{fnal_survey}; the depths of the shallow cubes were measured by hand
during the installation.  The cube positions are given in
Table~\ref{tab:cubes_positions}.

The time and charge measurements of each cube are calibrated with through-going
muons.  The charge gain calibration for each cube is performed by finding the
mean in the charge distributions of through-going muon events, assuming that
each muon is minimum ionizing.  The time calibration results in a time offset
such that the cube PMT hit matches the muon event time as measured by the main
detector.  The rate of muons passing through the muon tracker and one cube is
approximately half a Hertz, and the rate of muons stopping in the cubes is
approximately 100 per month.

\section{Detector Performance}
\subsection{Event Definition}
\label{sec:subevents}

\begin{figure}
\centerline{(a)\includegraphics[width=2.5in]{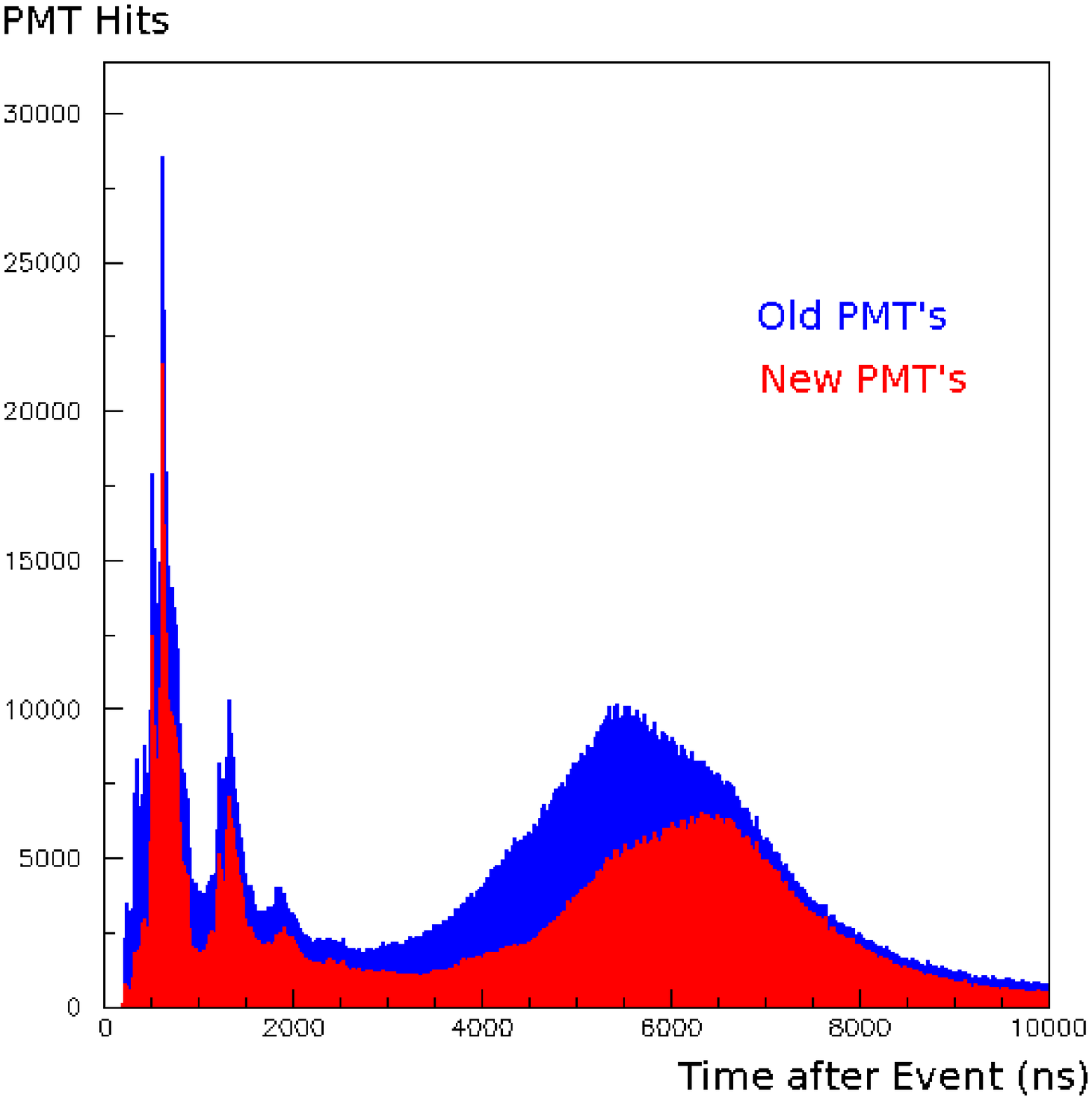}
\hfil (b)\includegraphics[width=2.5in]{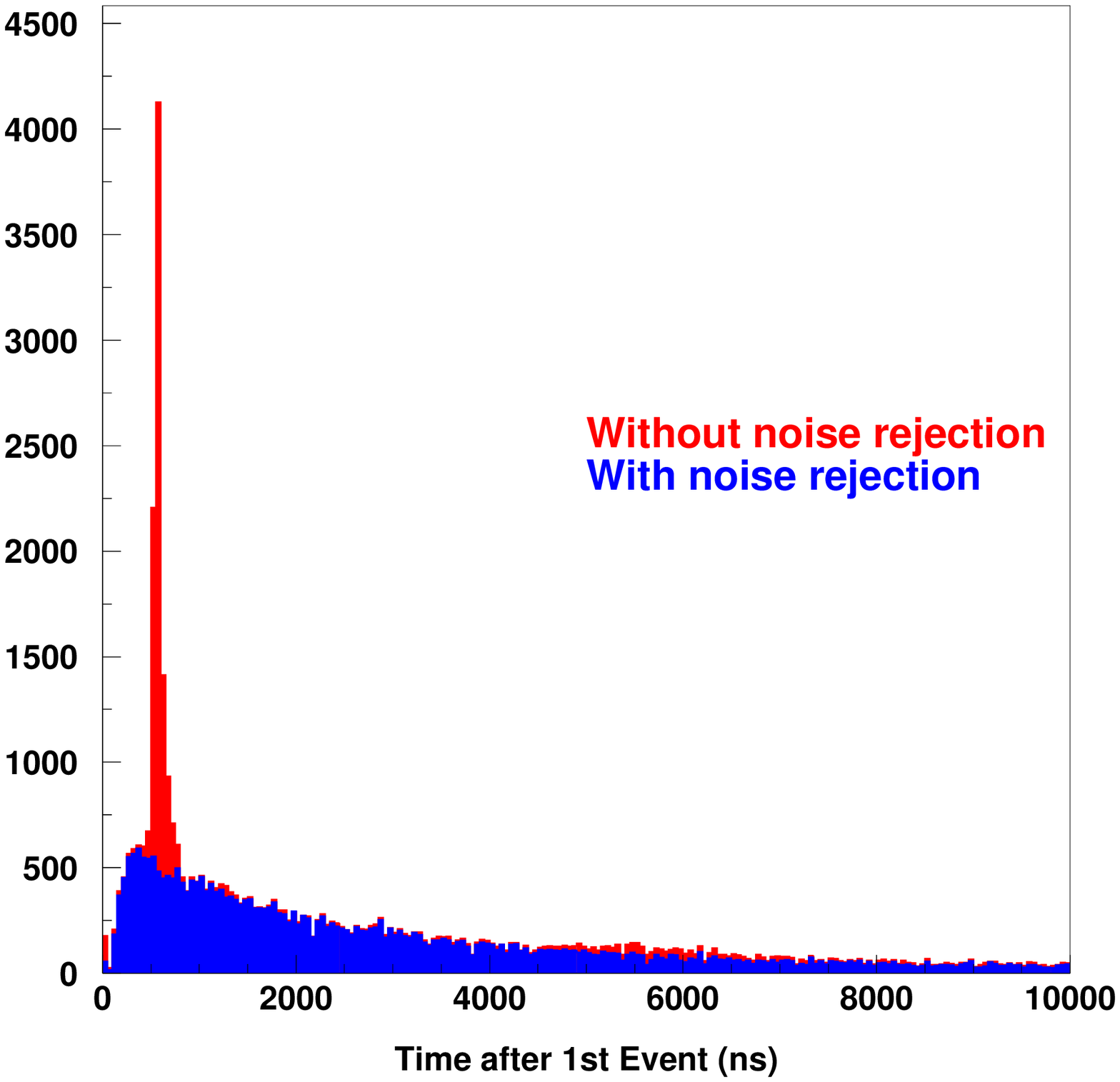}}
\caption{(a) Time distribution of PMT hits occurring after an energetic cosmic
ray event. Triggers that contained multiple hit clusters with high charge have
been excluded from this plot in order to remove muon decays and triggers with
more than one cosmic ray. (b) Time distribution of event clusters following
an energetic cosmic ray event with and without the noise rejection filter. The
filter removes the peaks due to coherent PMT noise without distorting the
underlying distribution expected from muon decays.}
\label{pmt_noise}
\end{figure}

As described in Sec.~\ref{sec:trigger}, a beam trigger records all PMT signals
that occurred in a 19.2~$\mu$s time window beginning approximately 5~$\mu$s
before the arrival of the neutrino beam regardless of whether or not there were
any interactions. Therefore the first step in any analysis is to isolate those
signals caused by interactions in the detector. This is done by looking for
clusters of PMT hits in time. A neutrino interaction can generate multiple time
clusters depending on the type of interaction. A $\nu_\mu$ charged current
quasi-elastic (CCQE) interaction will typically generate two clusters; one from
the muon produced in the initial reaction, followed by a delayed cluster from
the Michel electron produced by the muon decay at rest. CCQE interactions of
$\nu_e$'s will produce only single clusters. An event is defined as a trigger
containing at least one cluster and we refer to the individual clusters that
make up an event as subevents.

Clearly the number of subevents present in an event is an important clue to
determining what type of event it is. The algorithm used to find subevents must
therefore be both efficient and background-free. The one used in this experiment
starts off by defining a seed cluster as a group of at least 10 hits for which
the largest time difference between any two consecutive hits is less than 10~ns.
Once a seed cluster is found it is extended to include late hits provided there
are no time gaps of more the 20~ns and no more than two time gaps greater than
10~ns. This extension was required to pick up late PMT hits from scintillation
light that might otherwise have been wrongly indentified as a second subevent. 
Monte Carlo studies found no cases where this algorithm left out more than 10
PMT hits that should have been assigned to a subevent.

Events that generate a lot of light in the detector were found to also generate
a small amount of coherent noise in the PMT's that could produce fake subevents.
The noise hits have low charge and exhibit a distinctive time structure as shown
in Fig.~\ref{pmt_noise} (a). The clustering algorithm uses this time structure
to calculate the likelihood that any given PMT hit was due to noise. This is
then used to reject these fake subevents. Figure~\ref{pmt_noise} (b) shows the
effect of applying the noise rejection to hit clusters that occurred after an
energetic cosmic ray event.

Using Monte Carlo generated Michel decays and $\nu_e$ CCQE interactions, it was
determined that the algorithm's efficiency at finding electron subevents
increases from 99.5\% for 10~MeV/$c$ momentum electrons to 100\% for momenta
above 100~MeV/$c$.

\subsection{Muon Energy and Angle Resolution}

The muon energy and angular resolution of the detector are measured using
the cosmic muon calibration system described in Sec.~\ref{sec:cubes}.
The selection of cosmic muon events requires a clean reconstruction
using the muon tracker and exactly two subevents in the tank with
appropriate cuts on the number of PMT hits in each subevent.

We measure the range in oil using muons that pass through the muon
tracker and stop in the scintillation cubes; the measured range is
used to calculate the muon kinetic energy.  The stopping muon events
are selected by matching the times of tank subevents with cube hits,
and appropriate cuts on the charge of the cube hit.  The event sample is refined further by requiring
the reconstructed Michel electron position be near the cube position.

The two points obtained from the muon tracker and the position of the cube are
used in a fit to a straight line whose intersection with
the optical barrier of the detector gives the muon's entry point.  The
distance from this entry point to the cube's position gives the muon's
range.  From this, the energy is determined using a lookup
table~\cite{mu_range} for muon range in polyethylene with the measured
density of the MiniBooNE oil.

\begin{figure}
\centerline{\includegraphics[width=6.0in]{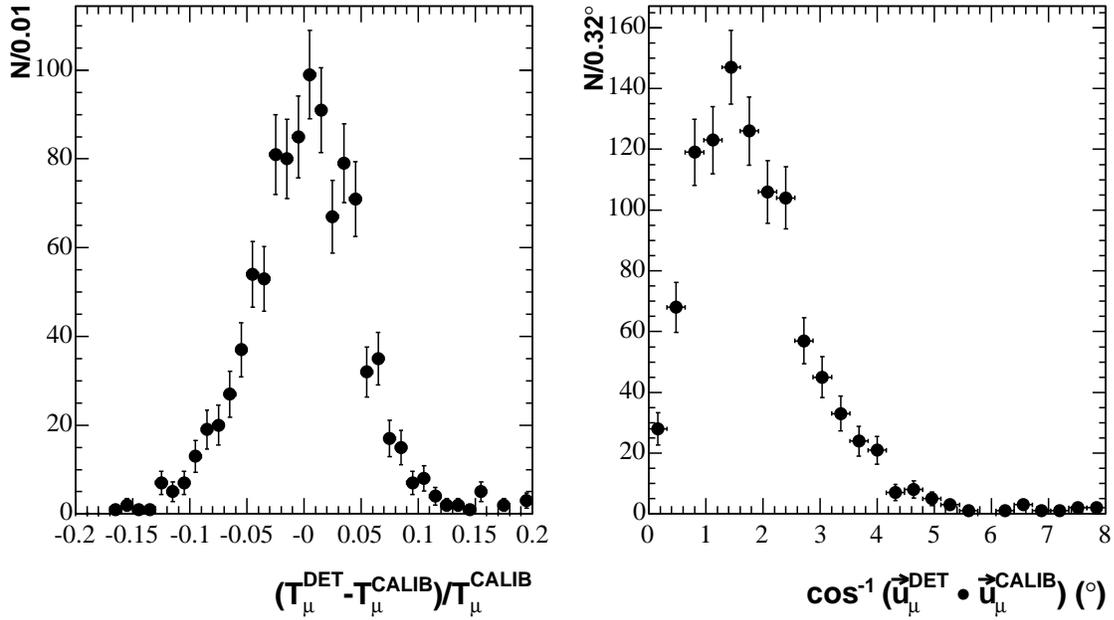}}
\caption{Energy and angle resolution for reconstruction of cosmic muons found
using the 2~m deep cube ($<T>$=407~MeV).}
\label{fig:cube1_resolutions}
\end{figure}

\begin{table}
\centerline{\begin{tabular}{|c|c|c|c|c|}\hline
$<T>$ (MeV) & Angular Resolution & Energy Resolution \\ \hline\hline     
95.  $\pm$ 4. &  5.4$^{\circ}$     &  12\%     \\
155. $\pm$ 5. &  3.2$^{\circ}$     &  7.0\%    \\
229. $\pm$ 7. &  2.2$^{\circ}$     &  7.5\%    \\
407. $\pm$ 9. &  1.4$^{\circ}$     &  4.6\%    \\
584. $\pm$ 9. &  1.1$^{\circ}$     &  4.2\%    \\
771. $\pm$ 9. &  1.0$^{\circ}$     &  3.4\%    \\ \hline
\end{tabular}}
\caption{The muon angle and kinetic energy resolution for each cube. The first
column is repeated from Table~\ref{tab:cubes_positions}}
\label{tab:cubes_resolutions}
\end{table}

The energy resolution is calculated from the fractional difference of kinetic
energies obtained from the detector event fitter~\cite{recon},
$T^{DET}_{\mu}$, and the cube track length calculation, $T^{CALIB}_{\mu}$.
Figure~\ref{fig:cube1_resolutions} shows the differences calculated for cosmic
muons stopping in the 2-m-deep cube.  The distribution is fitted to a Gaussian
function whose width gives the energy resolution.  The kinetic energy resolution
for each cube is shown in Table~\ref{tab:cubes_resolutions}; the energy
resolution thus extracted includes comparable contributions from $T^{DET}_{\mu}$
and $T^{CALIB}_{\mu}$ and is dominated by range straggling~\cite{sternheimer60}
and the sizes of the cubes themselves.

To study the angular resolution, we compare the direction vectors from
the detector event fitter, $\vec{u}^{DET}_{\mu}$ and the muon tracker,
$\vec{u}^{CALIB}_{\mu}$. The distribution of the angular difference between
these vectors for muon events stopping in the 2~m deep cube is shown in
Fig.~\ref{fig:cube1_resolutions}.  We fit the distribution to the
function $N(\theta) = c \cdot \theta \e^{-\theta^2/2\sigma_{\mu}^2}$,
under the assumption that the distribution represents the projection
of a symmetric two dimensional Gaussian distribution onto the radial
axis.  The width of the Gaussian function, $\sigma_{\mu}$, gives the
muon angular resolution.
The results for each cube are summarized in
Table~\ref{tab:cubes_resolutions}.

\subsection{Isolating Neutrino Interactions}
\label{sec:neutrinos}

\begin{figure}
\centerline{\includegraphics[width=6.0in]{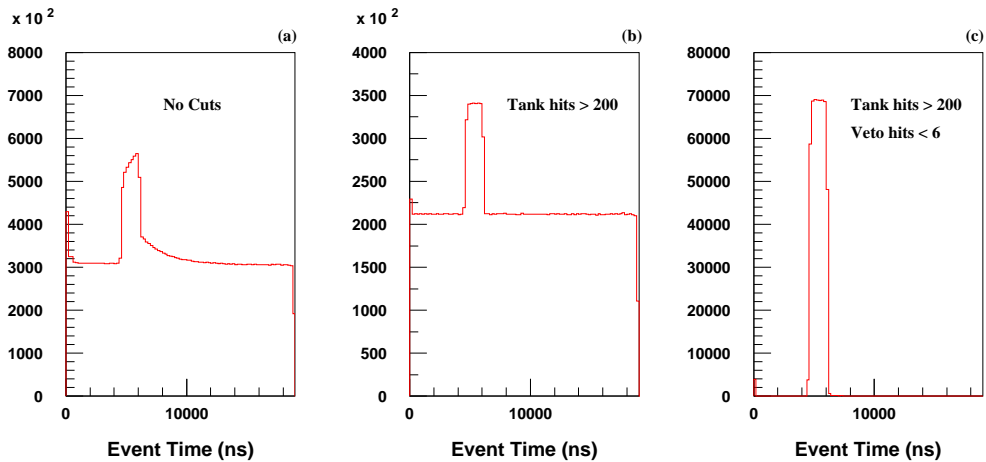}}
\caption{Time distributions of subevents within the 19.8 $\mu$-sec beam window:
a) with no cuts, b) subevents with more than 200 tank PMT hits, and c) subevents
with more than 200 tank PMT hits and fewer than 6 veto PMT hits.}
\label{event_time}
\end{figure}

Figure~\ref{event_time} shows time distributions of subevents within the beam
trigger windows. Figure~\ref{event_time} (a) contains all subevents without cuts
and clearly exhibits the three main components: a flat cosmic ray background,
neutrino beam interactions which have a flat distribution starting at about
4.5~$\mu$s and lasting for 1.6~$\mu$s, and Michel electrons from the decays of
muons produced in the neutrino interactions. The latter exhibit an exponential
fall off after the end of the beam spill and rise as 1 minus an exponential
during the spill. The exponential constants are characterized by the muon
lifetime. The pile up of events at the beginning of the trigger window is due to
cosmic rays that occurred just before the trigger and were still depositing
light in the tank when the readout started.

The first step in isolating subevents due to neutrino interactions is to remove
the low-energy ones produced by Michel electrons. This is done by rejecting subevents
with fewer than 200 PMT hits in the signal region of the main tank and the result
is shown in Fig.~\ref{event_time} (b). The remaining cosmic ray background is
removed by applying a cut on the number of PMT hits in the veto region of the
detector. Keeping only subevents with fewer than 6 veto hits reduces the cosmic
ray background by a factor of $10^4$ and leaves a very clean sample of beam
neutrino interactions as shown in Fig.~\ref{event_time} (c). Applying these cuts
to a high statistics sample of Strobe triggers, we obtain a veto efficiency of
99.987\% for cosmic ray events above the Michel electron energy cutoff.

\begin{figure}
\centerline{\includegraphics[width=4.0in]{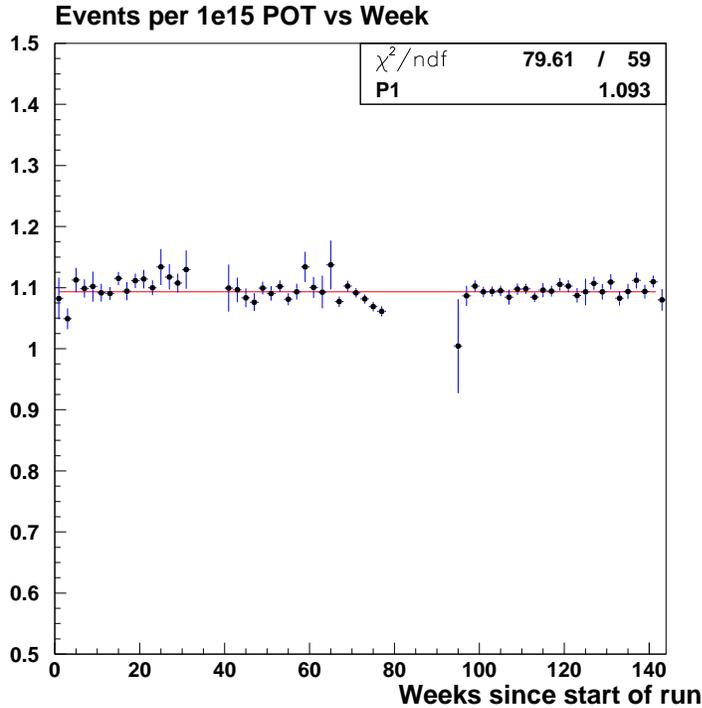}}
\caption{Neutrino interactions per beam proton as a function of time.}
\label{nu_per_pot}
\end{figure}

With a cleanly isolated sample of beam neutrino interactions, one can check the
stability of the detector and beam by monitoring the number of neutrino
interactions detected per proton on target (POT). In Fig.~\ref{nu_per_pot} this
ratio is plotted as a function of time over the 3 years in which MinBooNE
collected neutrino data. Statistical errors only are used in the flat line fit
to this data and the resulting $\chi^2$ per degree of freedom is consistent with
a less than 1\% week-to-week systematic error on the measurement of the protons
on target. There is no evidence of a long term trend that one might expect if,
for instance, the detector oil purity had been deteriorating. Instead, the data
indicate that the detector has performed well and remained stable since 2002
when it was first commissioned.

%

\section{Acknowledgments}

The authors acknowledge the support of Fermilab, the Department of Energy, and
the National Science Foundation. We thank Los Alamos National Laboratory (LANL)
for LDRD funding. We also acknowledge Dmitri Toptygin and Anna Pla for optical
measurements of the mineral oil. The MiniBooNE detector could not have been
built without the support provided by Fermilab, Los Alamos, and Princeton
University technicians. In particular Andy Lathrop from the Fermilab Mechanical
Department was crucial to the installation and maintenance of the various
utilities, including the nitrogen gas system. Camilo Espinoza and Neil Thompson
from LANL played a major role in the installation of the oil plumbing and
detector electronics. Ben Sapp, Shawn McKenney, and Nate Walbridge from LANL
were instrumental in the development and installation of the DAQ hardware and
software. The PMT testing and preparation owed much to the efforts
of Fermilab technicians, Sabina Aponte, Lyudmila Mokhov, and Galina Terechkina.
Hayes Lansford provided invaluable assistance with the tank scaffolding and
installation and finally, we would like to thank Princeton technicians, William
Groom, Stan Chidzik, and Robert Klemmer for their work on the fabrication and
installation of the PMT support structure.


\begin{thebibliography}{00}
\bibitem{lsnd}
A.~Aguilar {\em et~al.}, Phys.~Rev. {\bf D64}, 112007 (2001).

\bibitem{flux_paper}
A.~A.~Aguilar-Arevalo {\em et~al.}, ``The Neutrino Flux Prediction at MiniBoooNE'' (paper
submitted for publication: arXiv:0806.1449).

\bibitem{bonnie}B.T.~Fleming, L.~Bugel, E.~Hawker, V.~Sandberg, S.~Koutsoliotas, 
S.~McKenney and D.~Smith, IEEE~Trans.~Nucl.~Sci. {\bf 49}, 984 (2002).

\bibitem{hamamatsu} PMT efficiency data is from Hamamatsu Corp., D.~Lowell, private
communication, July 6, 2004.  It includes transmission through the glass, and this has
been corrected in the figure for the smaller reflection that occurs when the PMT is in oil.

\bibitem{Overview::Gladstone:2006}
S.~J.~Brice {\em et al.}, Nucl.~Instrum.~Methods, {\bf A562}, 97 (2006). 

\bibitem{Raaf:2002mt}
J.~L.~Raaf {\em et~al.}, IEEE~Trans.~Nucl.~Sci. {\bf 49}, 957 (2002).

\bibitem{Overview::Brown:2004}
B.~C.~Brown {\em et~al.},
IEEE Nuclear Science Symposium
Conference Record 1, 652 (2004).

\bibitem{Overview::LSND_NIM:1997}
C.~Athanassopoulos {\em et al.}, Nucl.~Instrum.~Methods, {\bf A388}, 149 (1997).

\bibitem{numi_alexis}
E.~Ables {\em et~al.}, FERMILAB-PROPOSAL-0875.

\bibitem{supernova_beacomb}
M.~K.~Sharp, J.~F.~Beacom, J.~A.~Formaggio, Phys.Rev. D66 (2002) 013012

\bibitem{ludox} Ludox is a registered trademark of W.~R.~Grace~Davison.

\bibitem{fnal_survey}
B.~O.~Oshinowo, Proceedings of 7th International Workshop on Accelerator Alignment (IWAA 2002), SPring-8, Japan, 11-14 Nov 2002, pp 021 (FERMILAB-CONF-02-425).

\bibitem{mu_range}
\texttt{http://pdg.lbl.gov/AtomicNuclearProperties}.

\bibitem{recon}
R.~B.~Patterson {\em et~al.}, ``The Extended-Track Event Reconstruction for
MiniBooNE'' (paper in preparation).

\bibitem{sternheimer60}
R.~M.~Sternheimer, Phys.~Rev. {\bf 117}, 485 (1960).

\end{thebibliography}
\end{document}